\documentclass[useAMS,usenatbib]{mn2e}

\usepackage[letterpaper,margin=0.635in]{geometry}
\usepackage[ansinew]{inputenc}
\usepackage[T1]{fontenc}
\usepackage{amsfonts}
\usepackage{amsmath}
\usepackage{enumerate}

\usepackage{enumitem}
\usepackage{graphicx}
\graphicspath{{./graphics/}}
\usepackage{caption} %
\usepackage{subcaption} %
\usepackage{epsfig}
\usepackage{color}
\usepackage[normalem]{ulem}
\usepackage{comment}
 
\usepackage{amssymb}
\usepackage{pdflscape}

\usepackage{booktabs}
\usepackage{multirow}

\usepackage[pagewise]{lineno}

\newcommand{\pz}{photo-$z$\ }
\newcommand{\pzs}{photo-$z$'s\ }

\newcommand{\ngmix}{\textsc{ngmix}\ }
\newcommand{\im}{\textsc{im3shape}\ }

\title[Galaxy bias from galaxy-galaxy lensing in DES-SV]{Galaxy bias from galaxy-galaxy lensing in the DES Science Verification Data}

\author[J. Prat et al.]{
\parbox{\textwidth}{
\Large
J.~Prat$^{1}$\thanks{Corresponding author: \texttt{\rm \texttt{jprat@ifae.es}}},
C.~S{\'a}nchez$^{1}$,
R.~Miquel$^{2,1}$,
J.~Kwan$^{3}$,
J.~Blazek$^{4}$,
C.~Bonnett$^{1}$,
A.~Amara$^{5}$,
S.~L.~Bridle$^{6}$,
J.~Clampitt$^{3}$,
M.~Crocce$^{7}$,
P.~Fosalba$^{7}$,
E.~Gaztanaga$^{7}$,
T.~Giannantonio$^{8,9}$,
W.~G.~Hartley$^{10,5}$,
M.~Jarvis$^{3}$,
N.~MacCrann$^{6}$,
W.J.~Percival$^{11}$,
A.~J.~Ross$^{4}$,
E.~Sheldon$^{12}$,
J.~Zuntz$^{6}$,
T. M. C.~Abbott$^{13}$,
F.~B.~Abdalla$^{10,14}$,
J.~Annis$^{15}$,
A.~Benoit-L{\'e}vy$^{16,10,17}$,
E.~Bertin$^{16,17}$,
D.~Brooks$^{10}$,
D.~L.~Burke$^{18,19}$,
A. Carnero Rosell$^{20,21}$,
M.~Carrasco~Kind$^{22,23}$,
J.~Carretero$^{7,1}$,
F.~J.~Castander$^{7}$,
L.~N.~da Costa$^{20,21}$,
D.~L.~DePoy$^{24}$,
S.~Desai$^{25}$,
H.~T.~Diehl$^{15}$,
P.~Doel$^{10}$,
T.~F.~Eifler$^{26}$,
A.~E.~Evrard$^{27,28}$,
A.~Fausti Neto$^{20}$,
B.~Flaugher$^{15}$,
J.~Frieman$^{15,29}$,
D.~W.~Gerdes$^{28}$,
D.~A.~Goldstein$^{30,31}$,
D.~Gruen$^{18,19,32}$,
R.~A.~Gruendl$^{22,23}$,
G.~Gutierrez$^{15}$,
K.~Honscheid$^{4,33}$,
D.~J.~James$^{13,34}$,
K.~Kuehn$^{35}$,
N.~Kuropatkin$^{15}$,
O.~Lahav$^{10}$,
M.~Lima$^{36,20}$,
J.~L.~Marshall$^{24}$,
P.~Melchior$^{37}$,
F.~Menanteau$^{22,23}$,
B.~Nord$^{15}$,
A.~A.~Plazas$^{26}$,
K.~Reil$^{19}$,
A.~K.~Romer$^{38}$,
A.~Roodman$^{18,19}$,
E.~Sanchez$^{39}$,
V.~Scarpine$^{15}$,
M.~Schubnell$^{28}$,
I.~Sevilla-Noarbe$^{39}$,
R.~C.~Smith$^{13}$,
M.~Soares-Santos$^{15}$,
F.~Sobreira$^{20,40}$,
E.~Suchyta$^{41}$,
M.~E.~C.~Swanson$^{23}$,
G.~Tarle$^{28}$,
D.~Thomas$^{11}$,
A.~R.~Walker$^{13}$
 \vspace{6mm}\\~\\
\parbox{\textwidth}{\centering \textsc{\Large(The DES Collaboration)} \\ 
\centering \textit{\small{Author affiliations are listed at the end of this paper}}\\ }}}

\begin{document}



\pagerange{\pageref{firstpage}--\pageref{lastpage}} \pubyear{0000}
\maketitle

\label{firstpage}


\begin{abstract}
We present a measurement of galaxy-galaxy lensing around a magnitude-limited ($i_{AB} < 22.5$) sample of galaxies from the Dark Energy Survey Science Verification (DES-SV) data. We split these lenses into three photometric-redshift bins from 0.2 to 0.8, and determine the product of the galaxy bias $b$ and cross-correlation coefficient between the galaxy and dark matter overdensity fields $r$ in each bin, using scales above 4 Mpc/$h$ comoving, where we find the linear bias model to be valid given our current uncertainties.
We compare our galaxy bias results from galaxy-galaxy lensing with those obtained from galaxy clustering \citep{Crocce2015} and CMB lensing \citep{Giannantonio2015} for the same sample of galaxies, and find our measurements to be in good agreement with those in \citet{Crocce2015}, while, in the lowest redshift bin ($z\sim0.3$), they show some tension with the findings in \citet{Giannantonio2015}. We measure $b\cdot r$ to be $0.87\pm 0.11$, $1.12 \pm 0.16$ and $1.24\pm 0.23$, respectively for the three redshift bins of width $\Delta z = 0.2$ in the range $0.2<z <0.8$, defined with the photometric-redshift algorithm BPZ. Using a different code to split the lens sample, TPZ, leads to changes in the measured biases at the 10-20\% level, but it does not alter the main conclusion of this work: when comparing with \citet{Crocce2015} we do not find strong evidence for a cross-correlation parameter significantly below one in this galaxy sample, except possibly at the lowest redshift bin ($z\sim 0.3$), where we find $r = 0.71 \pm 0.11$ when using TPZ, and $0.83 \pm 0.12$ with BPZ. 
\end{abstract}




\vspace*{3mm}

\section{Introduction}
\label{sec:intro}

Studying the large-scale structure of the Universe provides valuable information about its composition, origin and ultimate fate. Since most of the mass in the Universe is in the form of invisible dark matter, observations of galaxies must be used as a proxy to trace the dark matter on cosmological scales. However, galaxies are not perfect tracers of the underlying mass distribution, and thus, it is important to understand the relationship between the large-scale distribution of (mostly dark) matter and that of galaxies. Most of the cosmological information in the matter distribution can be encapsulated in the power spectrum of matter density fluctuations (all in the case of a Gaussian random field), $P_{\delta \delta}(k, z)$, as a function of wavenumber $k$ and redshift $z$. The power spectrum of the galaxy number density fluctuations, $P_{gg}(k, z)$, can then be related to the matter power spectrum as \citep{Kaiser1984, Bardeen1986}:
\begin{linenomath}
\begin{equation}\label{eq: intro}
P_{gg}(k, z) = b^2(k, z) \, P_{\delta \delta}(k, z)\ ,
\end{equation}
\end{linenomath}
where $b\,(k, z)$ is the so-called {\em galaxy bias} parameter, which is expected to be independent of $k$ at large separations (small enough $k$). It is therefore important to learn about the properties of the galaxy bias. 

One way to measure the galaxy bias is to use galaxy clustering, comparing the angular two-point correlation function of galaxies (essentially the Fourier transform of $P_{gg}$) with the theoretically-predicted matter two-point correlation function, to extract directly $b\,(z)$ at large-enough separation scales. Another way to probe the matter distribution is to use gravitational lensing (see \citealp{Bartelmann2001} for a review). A usual approach to measure gravitational lensing is to correlate some estimate of the lensing power with a tracer of the matter density field, such as galaxies. However, in this case, in the standard parametrization given by (\ref{eq: intro}), an additional factor appears to relate the matter power spectrum and the galaxy-matter cross-power spectrum $P_{g \delta}$ \citep{Dekel1999}: 
\begin{linenomath} \begin{equation}\label{eq: intro P_delta g = br P_delta delta}
P_{g \delta}(k, z)=  b\, (k, z) \, r\, (k, z)\, P_{\delta \delta} (k,z) ,
\end{equation}
\end{linenomath} 
where the cross-correlation parameter $r\,(k, z)$ \citep{Pen1998} connects not the amplitudes but the phases of the two distributions. If the distributions are completely correlated and thus the mapping between them is deterministic, then $r = 1$. On the other hand, if stochasticity and/or non-linearities are present in the relationship between the galaxy and matter distributions, then $r \neq 1$ \citep{Simon2007}. At large scales, however, $r$ is expected to be close to 1 \citep{Baldauf2010}. 

One possible way to probe the galaxy-matter cross-power spectrum is to use galaxy-CMB cross-correlations, first detected in \citet{Smith2007a}, where lensing maps of the Cosmic Microwave Background photons are cross-correlated with a density map of some foreground galaxies. Another possibility is to use galaxy-shear cross-correlations, or what is usually called galaxy-galaxy lensing \citep{Tyson1984, Brainerd1996}, which is the measurement of the tangential shear of background (source) galaxies around foreground (lens) galaxies. The amount of distortion in the shape of source galaxies is correlated with the amount of mass causing the light to curve. Galaxy-galaxy lensing at large scales has been used to probe cosmology, for instance in \citet{Mandelbaum2013} and in \citet{Kwan2016}, and at smaller scales to learn about the dark matter haloes, as in \citet{Sheldon2004}, \citet{Velander2014} and \citet{Hudson2015}.

The Dark Energy Survey (DES) is a large imaging survey that is in the process of mapping 5000 sq.~deg.~of the southern sky to a depth of $i_{AB}\sim24$ in five optical and near-infrared bands ($grizY$) during five seasons that started in August 2013. Before that, a Science Verification (SV) period of observations took place between November 2012 and February 2013 that provided science-quality imaging for almost 200 sq.~deg.~at the nominal depth of the survey. As described above, there are many ways to obtain information on $b\,(k, z)$, and several have already been attempted with this DES-SV data set. In \citet{Crocce2015} (henceforth Cr16), galaxy clustering measurements were performed to obtain the galaxy bias. The results, depicted in fig.~11 in Cr16, show a moderate increase of the galaxy bias with redshift, an increase that is expected based on numerical simulations \citep{Gaztanaga2012}, and also observed in other studies such as \citet{Coupon2012} from CFHTLS measurements. In \citet{Giannantonio2015} (henceforth G16) galaxy-CMB lensing cross-correlations of the same foreground galaxy sample as in Cr16 were presented, providing another measurement of the relationship between the mass and galaxy distributions. The results in G16, displayed in their fig.~21, show a moderate tension with those in Cr16, of $\sim 2\sigma$ using the full galaxy sample at $0.2 < z < 1.2$, and particularly at the lowest redshift, where the tension is $\sim 3\sigma$. Since the two galaxy samples are identical, the most straightforward way to reconcile the two measurements within the standard $\Lambda$CDM cosmological model is by assuming that $r$ differs significantly from 1 ($r \lesssim 0.6$) at redshift $z \sim 0.3$, a somewhat unexpected result.

In this work, we provide a third probe to measure the galaxy bias, using galaxy-galaxy lensing on the same foreground galaxy sample as the one used in Cr16 and G16, so that we can readily compare our results and shed light on the apparent tension mentioned above. The background set of galaxies we use is that introduced in~\citet{Jarvis2015}, which was used in previous DES weak lensing analyses \citep{Baxter2016, Becker2015, Kacprzak2016, Kwan2016, Sanchez2016, TheDarkEnergySurveyCollaboration2015}. Particularly, \citet{Clampitt2016} performed a series of shear tests using galaxy-galaxy lensing with the DES redMaGiC sample \citep{Rozo2015} as lenses. Note that the background (source) galaxy sample only serves to illuminate the foreground (lens) sample, which is the one we will gain knowledge of. An advantage of this method is that, since it involves the cross-correlation between source galaxy shapes and lens galaxy density, it is, at least at first order, insensitive to those additive systematic effects that affect only one of these two galaxy samples, such as additive shear biases.

Along similar lines, \citet{Chang2016} used the ratio between the (foreground) galaxy density maps and the mass maps obtained from weak lensing in DES-SV to determine the galaxy bias parameter for the same galaxy sample as in Cr16 and G16. The approach used in \citet{Chang2016} has the advantage of being weakly dependent on the assumed cosmological parameters, such as the amplitude of the power spectrum of matter fluctuations, $\sigma_8$, but, on the other hand, in the relatively small DES-SV sample, its statistical power is somewhat limited. \citet{Chang2016} assumed $r=1$, and, as shown in their fig.~6, their results are generally more in agreement with those in Cr16, although the errors are large. The measurements presented in this work are sensitive to the product $b\cdot r$ and therefore can help resolve the apparent discrepancy between the results in G16 (that measure $b\cdot r$ as well) and Cr16 (that measure $b$).

Since the main goal of the paper is to compare with the galaxy bias results in Cr16 and G16, the same lens galaxy sample is used, despite its limited resolution. Then, our lens and source samples, defined in Sec.~\ref{sec:description of the data}, significantly overlap in redshift. Other studies of galaxy-galaxy lensing \citep{Nakajima2012,Hudson2015} have chosen to eliminate pairs of lens-source galaxies that are close in estimated redshift. We instead model the overlap in the computation of the predicted signal, which relies on the calibrated redshift distributions for lenses and sources, as described in \citet{Sanchez2014}, Cr16 and \citet{Bonnett2015}. In the DES-SV papers that use galaxy-galaxy lensing to obtain cosmological results \citep{Clampitt2016,Kwan2016}, an alternative lens sample composed of luminous red galaxies selected using the redMaGiC algorithm \citep{Rozo2015} was instead used, with very precise photometric redshifts ($\sigma(z) \simeq 0.02$). A similar redMaGiC lens sample has been used for the DES Year 1 cosmological analysis \citep{Prat2017}.

The outline of this paper is as follows. First, in Sec.~\ref{sec:method} we explain the theory and the method employed to measure the galaxy bias using galaxy-galaxy lensing; next, in Sec.~\ref{sec:description of the data} the data that we use are described; then, in Sec.~\ref{sec:measurement methodology}, we present the methodology used to do our measurements and obtain the results, which are presented in Sec.~\ref{sec:results}; in Sec.~\ref{sec:systematics} we discuss the possible implications potential systematics might have on our measurements; finally, in Sec.~\ref{sec:discussion} we further discuss our results comparing them to previous work and conclude.
\section{Theory and method}
\label{sec:method}

Our goal is to measure the galaxy bias using galaxy-galaxy lensing, which measures the effect some foreground mass distribution traced by galaxies (lenses) has on the shapes that we observe of some other background galaxies (sources). This small distortion on the shape of the galaxy image is referred to as cosmic shear. The main observable of galaxy-galaxy lensing is the tangential shear, which can be expressed as a function of the cross-power spectrum  $\mathcal{C}_{g \kappa}$:
\begin{linenomath} 
\begin{equation}
\begin{split}\label{eq: gamma_t theory}
\gamma_t  (\theta) = \int \frac{\text{d}\ell}{2\pi} \, \ell \, J_2(\theta \ell )\,  \mathcal{C}_{g \kappa}(\ell),
\end{split}
\end{equation}
\end{linenomath} 
where $\mathcal{C}_{g \kappa}$ is the projection along the line of sight of the $\text{3-D}$ galaxy-matter cross-power spectrum $P_{g \delta} $ and $J_2$ is the second order Bessel function. $\ell$ is the multipole moment, which is the 2-D analogous of the 3-D wavenumber $k$, and both can be easily related using the Limber approximation $k=\ell/\chi$ \citep{Limber1953}. Therefore, $\mathcal{C}_{g	\kappa }$ can be expressed as the line of sight projection of the 3-D cross-power spectrum $P_{g \delta} $ in the following way:
\begin{linenomath} \begin{equation}\label{eq: C_gk}
\mathcal{C}_{g \kappa}(\ell) = \frac{3}{2} \, \Omega_m  \left( \frac{H_0}{c}\right)^2 
 \int \text{d}\chi \, \frac{g(\chi)\,N_l(\chi)}{a(\chi)\, \chi } \, P_{g \delta} \left( k= \frac{\ell}{\chi }, \, \chi \right), 
 \end{equation}
\end{linenomath} 
where $\chi$ is the comoving distance to a lens galaxy, $N_l(\chi)$ is the distribution of the lens sample, $a(\chi)$ is the scale factor and $g(\chi)$ is the lens efficiency factor:
\begin{linenomath} \begin{equation}\label{eq: lens efficiency factor}
g(\chi) = \int_{\chi}^{\chi_h} \text{d}\chi '\, N_s (\chi ') \, \frac{\chi ' - \chi}{\chi '},
\end{equation}
\end{linenomath} 
where $\chi_h$ is the comoving horizon distance and $N_s(\chi ')$ the distribution of the source sample in comoving distance. For this analysis, we measured the redshift distribution for both the lens and the source sample $N'(z)$, which we then converted to the distribution in comoving distance using the relation $N'(z)\text{d}z = N(\chi)\text{d}\chi$. 

On the other hand, the cross-power spectrum $P_{g \delta} $ can be related to the matter power spectrum $P_{\delta \delta}$ through the galaxy bias $b$ and the cross-correlation coefficient between matter and galaxy fluctuations $r$:
\begin{linenomath} \begin{equation}\label{eq: P_delta g = br P_delta delta}
P_{g \delta}(k, \chi)=  b\, (k, \chi) \, r\, (k, \chi)\, P_{\delta \delta} (k, \chi) ,
\end{equation}
\end{linenomath} 
with
\begin{linenomath} \begin{equation}
b\,(k, \chi) = \sqrt{\frac{P_{gg}(k, \chi)}{P_{\delta \delta}(k, \chi) }} , 
\end{equation}
\end{linenomath} 
so that
\begin{linenomath} \begin{equation} \label{eq: r}
r\,(k, \chi) = \frac{P_{g\delta}(k, \chi)}{\sqrt{P_{\delta \delta}(k, \chi) \, P_{gg}(k, \chi)}} .
\end{equation}
\end{linenomath} 

Then, combining (\ref{eq: P_delta g = br P_delta delta}) with (\ref{eq: C_gk}), it is possible to express $\mathcal{C}_{g \kappa}$ as a function of the product of the galaxy bias times the cross-correlation coefficient $b\cdot r$. Furthermore, (\ref{eq: gamma_t theory}) and (\ref{eq: C_gk}) can be combined to relate the tangential shear $\gamma_t (\theta)$ to the factor $b \cdot r $:

\begin{linenomath} \begin{equation}
\begin{split}\label{eq: gamma_t explicit}
\gamma_t  (\theta) &=  \frac{3}{2} \, \Omega_m  \left( \frac{H_0}{c}\right)^2  \int \frac{\text{d}\ell}{2\pi} \, \ell \, J_2(\theta \ell )\, 
 \int \text{d}\chi \,\left[ \frac{g(\chi)\,N_l(\chi)}{a(\chi)\, \chi } \, \right. \\
 &\left.  b\, (k = \frac{\ell}{\chi }, \chi) \, r\, (k = \frac{\ell}{\chi }, \chi)\, P_{\delta \delta} \left( k= \frac{\ell}{\chi }, \, \chi \right)\right].
\end{split}
\end{equation}
\end{linenomath} 

However, both the galaxy bias and the cross-correlation factor depend on the scale and on the comoving distance to the lens galaxy $\chi$, or similarly, on redshift.
Then, assuming $b\cdot r$  is redshift and scale independent in the lens sample considered, the factor $b\cdot r$  can be taken out of the integrals along the line of sight and over the scales in (9). In this case, $\gamma_t $ is directly proportional to $b\cdot r$, which in reality is an effective average over the redshift range of the given bin and the scales considered in the measurement.

For instance, these are the circumstances on large scales -- larger than a few times the typical size of a dark matter halo \citep{Mandelbaum2013} -- where the galaxy bias tends to a constant value and we can use the linear bias approximation. The cross-correlation coefficient is also expected to be scale independent at large scales, approaching unity \citep{Baldauf2010}. The dependence on $\chi$, or equivalently redshift, can be avoided using narrow-enough redshift bins and assuming the galaxy bias does not evolve within them.

Hence, the factor $b\cdot r$ of a lens galaxy sample can be measured by comparing the predicted or modelled tangential shear using (\ref{eq: gamma_t explicit}), with $b\cdot r = 1$, to the measured tangential shear around the lens galaxy sample. We compute the non-linear power spectrum with \texttt{Halofit} \citep{Smith2003, Takahashi2012} using \texttt{CosmoSIS}\footnote{\texttt{https://bitbucket.org/joezuntz/cosmosis/wiki/Home}} \citep{Zuntz2015}.
 
To model the tangential shear, we assume a fiducial flat $\Lambda$CDM+$\nu$ (1 massive neutrino) cosmological model based on the Planck 2013 + WMAP polarization + highL(ACT/SPT) + BAO best-fit parameters \citep{Ade2014}, consistently with Cr16 and G16: $\Omega_m = 0.307$,  $\Omega_{\nu} = 0.00139$, $\Omega_b = 0.0483$, $\sigma_8 = 0.829$, $n_s = 0.961$, $\tau = 0.0952$ and $h= 0.678$.

\section{Description of the data}
\label{sec:description of the data}

DES \citep{DES2016} is a photometric survey that will cover $\sim 5000$ sq. deg. of the southern sky by the end of its five year observation program using the Dark Energy Camera (DECam, \citealp{Flaugher2015}), a 570-Megapixel digital camera mounted at the prime focus of the Blanco \mbox{4-meter} telescope at Cerro Tololo Inter-American Observatory. Five filters are used (\textit{grizY}) with a nominal limiting magnitude $i_{AB}\simeq 24 $ and with a typical exposure time of 90 sec for \textit{griz} and 45 sec for \textit{Y}. The \textit{Y} band is not used for the measurements in this paper.

DES officially began in August 2013. Prior to the main survey, the Science Verification (SV) data were taken, from November 2012 to February 2013. During the SV period, $\simeq 200$ sq. deg. of the sky were imaged to the nominal DES depth, which produced a usable catalog for early science results. The region used in this work is the largest contiguous area in the SV footprint, contained in the South Pole Telescope East (SPT-E) observing region with $60^\circ < \text{RA} < 95^\circ$ and $-61^\circ< \text{Dec} < -40^\circ$, which covers $163$ sq. deg.

The most numerous catalog of reliable objects in DES-SV is the SVA1 Gold Catalog\footnote{Publicly available at \texttt{https://des.ncsa.illinois.edu/releases/sva1}}, which excludes objects that are known to be problematic in some way, because of, for instance, failed observations or imaging artefacts. It is generated by applying the cuts and conditions described in \citet{Jarvis2015}. The SPT-E region of the Gold Catalog covers $ 148$ sq. deg. of the sky.

\subsection{The lenses: The Benchmark sample}\label{subsec: the lenses}

The foreground catalog for the galaxy-galaxy lensing measurements in this work is the Benchmark sample, which is a cleaner subsample of the Gold Catalog. The Benchmark sample was first introduced in Cr16 to perform galaxy clustering measurements, and it was used in G16 to perform measurements of CMB lensing around foreground galaxies. From the SPT-E region of the Gold Catalog, the Benchmark sample is selected by applying the following selection criteria:

\begin{itemize}[leftmargin=.2in]\setlength\itemsep{0.1in}
\item $\text{Dec} > -60^\circ$: Conservative cut to remove any possible contamination from the LMC\footnote{Note that in G16 a slightly different cut of $\text{Dec} > -61^\circ$ was applied.}. 
\item $18 < i < 22.5$: Magnitude cut in the \mbox{$i$-band}, where $i$ refers to \texttt{SExtractor}'s \texttt{MAG\_AUTO}  \citep{Bertin1996}. 
\item $-1 < g -r < 3$, $-1 < r-i < 2$, $-1 < i-z < 2$: Color cuts to remove outliers in color space. In this case, the magnitude used is \texttt{MAG\_DETMODEL} since it produces more accurate colors than \texttt{MAG\_AUTO}. 
\item \texttt{WAVG\_SPREAD\_MODEL} $> 0.003$: star-galaxy separation cut. \texttt{SPREAD\_MODEL} is a \texttt{SExtractor} parameter that measures the light concentration of an object \citep{Desai2012}. \texttt{WAVG\_SPREAD\_MODEL} is the weighted average of the \texttt{SPREAD\_MODEL} values for all single epoch images used to coadd one object.
\item More conservative cut to remove defective objects than the one applied in the Gold Catalog\footnote{\texttt{Badflag} $\leq 1$.}. 
\end{itemize}

Furthermore, a mask which ensures the completeness of the sample is applied. Only regions deeper than $i=22.5$ are included (Cr16), providing a catalog with $2,333,314$ objects remaining, covering 116.2 sq. deg. 

\subsection{The Sources: Shear Catalogs} \label{subsec: The sources: shear catalogs}

The source catalogs for this work are the SV shear catalogs \ngmix and \im, which have been produced for a subset of objects of the DES-SV Gold Catalog in the SPT-E region. \ngmix and \im are two independent shear pipelines both based on model-fitting algorithms, which are discussed in detail in \citet{Jarvis2015}. Throughout this work, the \ngmix shear catalog is used as the fiducial source catalog, since it has  a larger raw galaxy number density, 6.9 arcmin$^{-2}$, as opposed to 4.2 arcmin$^{-2}$ for \textsc{im3shape} \citep{Jarvis2015}. 

The main features of both shear pipelines are described below:

\begin{itemize}[leftmargin=.2in] \setlength\itemsep{0.1in}
\item {\bfseries{\scshape ngmix}:} The \ngmix shear pipeline \citep{Sheldon2014} for the DES-SV catalogs uses an exponential disk model for the galaxy, which is fit simultaneously in the $riz$ bands. To estimate the ellipticity, the \textsc{lensfit} algorithm \citep{Miller2007} is used. The \textsc{lensfit} method requires a prior on the ellipticity distribution $p\,(e)$, taken from the galaxies in the COSMOS catalog \citep{Koekemoer2007}. 

\item {\bfseries{\scshape im3shape}:} \im is based on the algorithm in \citet{Zuntz2013}, modified according to \citet{Jarvis2015}. It performs a maximum likelihood fit using a bulge-or-disk galaxy model to estimate the ellipticity of a galaxy, that is, it fits de Vaucouleurs bulge and exponential disk components to galaxy images in the $r$ band (in the case of SV data). 
\end{itemize} 

A weight factor $\omega$ related to the uncertainty in the measurement of the galaxy shape is assigned to each object in the following way:
\begin{linenomath}
\begin{equation}\label{eq: weights}
\omega =\frac{1}{\sigma_\text{SN}^2+ \sigma_e^2},
\end{equation}
\end{linenomath}
where $\sigma_\text{SN}$ represents the shape noise per component -- the standard deviation of the intrinsic ellipticities -- and $\sigma_{e}$ the measurement uncertainty, estimated in different ways for both shear catalogs \citep{Jarvis2015}.  

Also, both for \im and \textsc{ngmix}, the raw values in the catalogs are intrinsically biased estimators of the shear in the presence of noise (\citealt{Refregier2012, Kacprzak2012}). We correct for this noise bias in the shear measurement as explained in \citet{Jarvis2015}.

\subsection{Photometric redshifts} \label{subsec:photoz codes}

In this work, point estimates of the redshift are necessary to divide the galaxies into redshift bins (see Sec.~\ref{subsec:redshift bins}) and therefore allow for a tomographic study of the galaxy bias. On the other hand, the redshift distributions $N(z)_{l,s}$ of lenses and sources are also needed to model the tangential shear and consequently to measure the galaxy bias (see Sec.~\ref{subsec:N(z)}). For this reason, it is advantageous to estimate the whole redshift probability density function $P(z)$ for each galaxy, which can then be stacked for a collection of galaxies to obtain $N(z)$. Then, point estimates of the redshift for each galaxy are obtained by taking the mean of each $P(z)$.

Since DES is a photometric survey, redshifts are measured using photometry. There are two main approaches to estimate photometric redshifts or photo-$z$'s: template-based and training-based methods. Template-based methods use galaxy templates to match the measured photometry to the best fit template. On the other hand, training-based methods rely on machine learning algorithms trained on spectroscopic samples. The SV area was chosen to overlap with several deep spectroscopic surveys, such as VVDS \citep{Fevre2004}, ACES \citep{Cooper2012} and zCOSMOS \citep{Lilly2007} to be able to calibrate the \pzs. This is not an easy task and currently there exist several photometric redshift codes with different performances that were discussed in detail in \citet{Sanchez2014}, for DES-SV, and in \citet{Bonnett2015}, for DES-SV shear catalogs. In \citet{Bonnett2015}, the four best performing \pz codes according to \citet{Sanchez2014} were studied: TPZ \citep{CarrascoKind2013}, BPZ \citep{Benitez2000a, Coe2006}, SkyNet  \citep{Graff2014,Bonnett2015a} and ANNz2 \citep{Sadeh2015}, which are the ones that are employed in this work to estimate the $N(z)$ of the shear catalog. On the other hand, in both Cr16 and G16 only two \pz codes are used for the $N(z)$ of the Benchmark sample: TPZ and BPZ, which we will adopt in this work as well. A brief description of each of these \pz codes is given below:  

\begin{itemize}[leftmargin=.2in] \setlength\itemsep{0.1in}

\item {\textbf{BPZ}} \citep{Benitez2000a, Coe2006} is a template-based method  that provides the probability density distribution $p(z|m_i, \sigma_i)$ that a galaxy with magnitudes in each band $m_i	\pm \sigma_i$ is at redshift $z$. 

\item {\textbf{TPZ}} \citep{CarrascoKind2013} is a training-based code based on prediction trees and random forest algorithms. 

\item {\textbf{SkyNet}} \citep{Graff2014, Bonnett2015a} is a training-based method using a neural network algorithm to classify galaxies in classes, in this case redshift bins. 

\item {\textbf{ANN$\boldsymbol{z}$2}} \citep{Sadeh2015} is the updated version of ANN$z$ (Artificial Neural Network). It is a training-based method which relies on Artificial Neural Networks (ANNs), and in the updated version ANN$z$2, also on Boosted Decision Trees (BDTs) and K-Nearest Neighbours (KNNs), as implemented in the TMVA package \citep{Hoecker2007}. 
\end{itemize} 

\subsection{Redshift bins} \label{subsec:redshift bins}

\begin{table}
\centering
\begin{tabular}{lll} \toprule 
& Lens redshift bins &  Source redshift bins\\
\midrule 
\multirow{2}{*}{1st} & \multicolumn{1}{l}{$0.2 \leq z_{\text{TPZ}} < 0.4$  }   & \multirow{2}{*}{$0.55 < z_{\text{SkyNet}} < 1.3$} \\ 
&\multicolumn{1}{l}{$0.2 \leq z_{\text{BPZ}} < 0.4$ } \\ [0.2cm] 

\multirow{2}{*}{2nd} & \multicolumn{1}{l}{$0.4 \leq z_{\text{TPZ}} < 0.6$ }   & \multirow{2}{*}{$0.55 < z_{\text{SkyNet}} < 1.3$} \\ 
& \multicolumn{1}{l}{$0.4 \leq z_{\text{BPZ}} < 0.6$ } \\  [0.2cm] 

\multirow{2}{*}{3rd} & \multicolumn{1}{l}{$0.6 \leq z_{\text{TPZ}} < 0.8$ }   & \multirow{2}{*}{$0.83 < z_{\text{SkyNet}} < 1.3$} \\ 
&\multicolumn{1}{l}{$0.6 \leq z_{\text{BPZ}} < 0.8$ } \\  
\bottomrule
\end{tabular}
\caption{ \label{tab:redshift bins}Definition of the lens and source redshift bins used throughout the paper. The lens redshift bins are identical to the ones in Cr16 and G16, defined with BPZ and TPZ as well. The source redshift bins are the same studied in \citet{Bonnett2015} and used in other DES-SV weak lensing analyses, where SkyNet is used to define the bins. $z_{\text{TPZ}}$, $z_{\text{BPZ}}$ and $z_{\text{SkyNet}}$ stand for the mean of the \pz probability density function for each galaxy determined with each code.}
\end{table}

\begin{table}
\centering
\begin{tabular}{ccc}\toprule 
 Lens redshift bin & $N_\text{TPZ}$ &  $N_\text{BPZ}$  \\
\midrule 
$0.2 \leq z_l < 0.4$ & 398,658 & 551,257 \\
$0.4 \leq z_l < 0.6$ & 617,789 & 647,010 \\
$0.6 \leq z_l < 0.8$ & 586,298 & 494,469 \\
\bottomrule
\end{tabular}

\caption{Number of galaxies in each redshift bin after the veto mask (see Sec.~\ref{subsec: mask}) has been applied. \label{tab:number of galaxies}}
\end{table}

\begin{figure}
    \centering
    \includegraphics[width=0.48\textwidth]{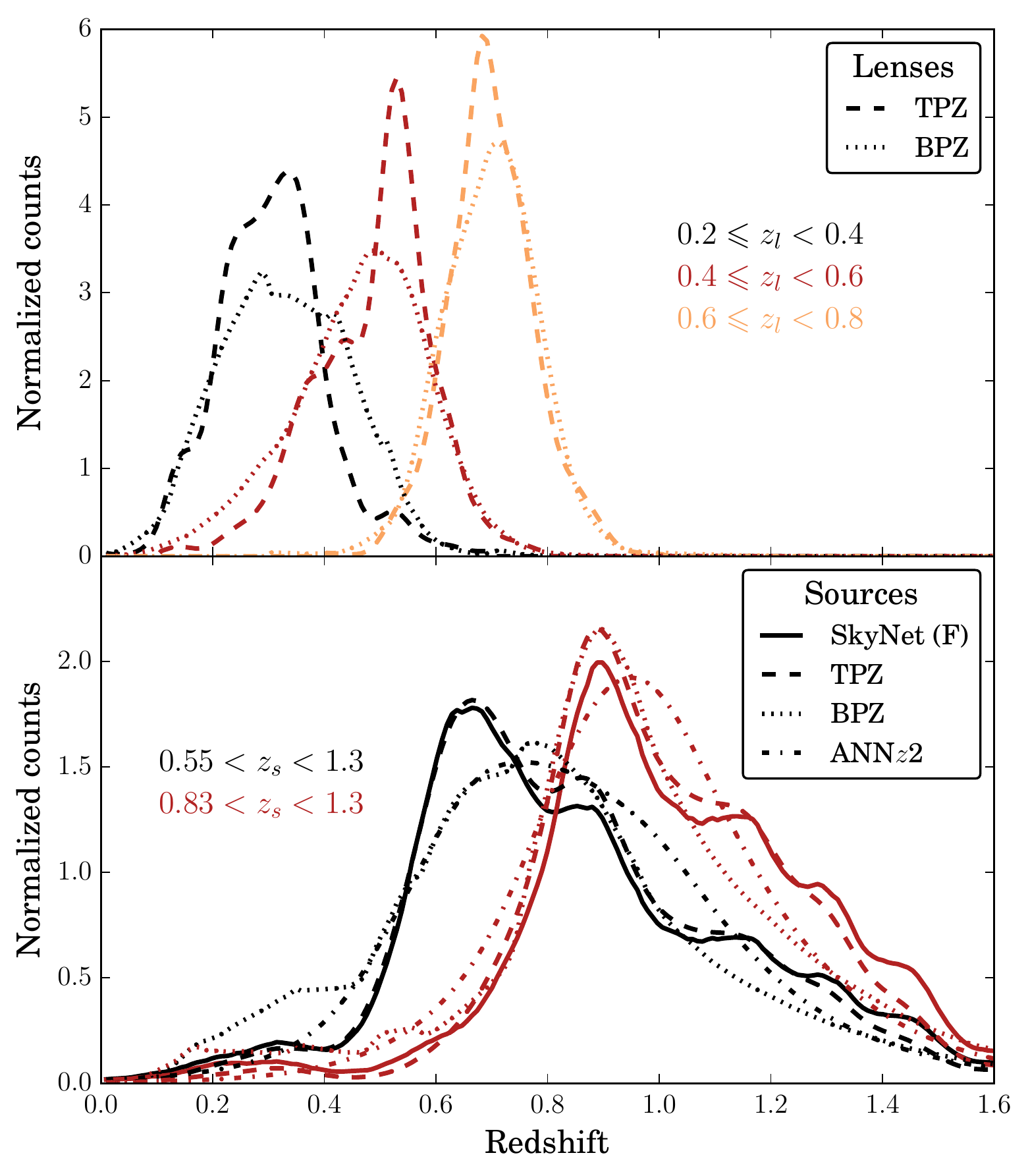}
    \vspace*{-5mm}
    \caption{\textit{Top panel:} Normalized counts (or normalized $N(z)$'s) of the foreground redshift distribution of lens galaxies using two \pz codes: TPZ and BPZ, also used in each case to define the galaxies that belong to each bin (as described in Table \ref{tab:redshift bins}). \textit{Bottom panel:} Normalized counts of the background redshift distribution of source galaxies from \ngmix using the following four \pz codes: SkyNet, TPZ, BPZ and ANN$z$2, for both source redshift bins. We choose SkyNet to be the fiducial \pz code to estimate the $N(z)$ of the source galaxies.
     }\label{fig: nofz}
\end{figure}

\subsubsection{Lens redshift bins}

The lens sample is divided into three \pz bins of width $\Delta z = 0.2$, from $z=0.2 $ to $z = 0.8$, as in Cr16 and G16, and as shown in Table~\ref{tab:redshift bins}. The objects are classified into these lens redshift bins using the mean of the \pz probability density function determined with either BPZ or TPZ, like in Cr16 and in G16, for comparison. The \pz precision $\sigma_{68}$ (half the width of the distribution, centered at the median, where 68\% of the data are enclosed) was measured to be $\sim 0.1$ for BPZ and $\sim 0.08$ for TPZ in \citet{Sanchez2014}. Therefore, it is suitable to use redshift bins of width $\Delta z = 0.2$, being approximately twice the \pz precision $\sigma_{68}$. The number of galaxies in each redshift bin after the veto mask (see Sec.~\ref{subsec: mask}) is applied is given in Table~\ref{tab:number of galaxies}.

In Cr16 and in G16 two additional high redshift bins were used, from $z = 0.8$ to $z = 1$ and from  $z = 1$ to  $z = 1.2$. However, both are omitted in this work, since there is not enough galaxy-galaxy lensing signal as the number of source galaxies at $z>1$ is limited.

\subsubsection{Source redshift bins}

For this work, the two high-redshift source redshift bins studied in \citet{Bonnett2015} are adopted, to be consistent with other DES-SV analyses \citep{Becker2015,TheDarkEnergySurveyCollaboration2015}. These are defined from $z = 0.55$ to $z=0.83$ and from $z=0.83$ to $z=1.3$, using the mean of the SkyNet probability density function as a point estimate of the redshift to define the bins. For the two lower redshift bins of the lenses, we use as source redshift bin the combination of both bins, from $z=0.55$ to $z=1.3$ to increase the number of sources and thus the signal to noise, while for the third lens bin, only the higher source redshift bin, from $z=0.83$ to $z=1.3$ is used, as shown in Table~\ref{tab:redshift bins}. 

\subsection{Veto Mask}\label{subsec: mask}

Besides the depth mask described in Sec.~\ref{subsec: the lenses}, a veto mask characterized in Cr16, removing the areas most affected by systematics, is also applied in some of the redshift bins. Using the maps of potential sources of systematics presented in \citet{Leistedt2015}, Cr16 studied the relationship between the galaxy density and several potential systematics, such as seeing or airmass. In some cases, they found the galaxy density to drop from its mean value in areas with extreme systematics contamination. Then, the regions corresponding to these systematics values are removed, hence defining a veto mask. In detail, they found the seeing to be the main quantity influencing the galaxy density in this manner, differently for the various redshift bins. For the lowest redshift bin $0.2 \leq z_l < 0.4$, 19.5\% and 9.7\% of the galaxies are removed, for the BPZ and TPZ redshift bins, respectively. The veto mask for the $0.4\leq z_l <0.6$ is the same for both the BPZ and TPZ redshift bins, removing 14.8\% and 14.4\% of the galaxies, respectively. On the other hand, the highest redshift bin $0.6\leq z_l < 0.8$ is found to be less affected by systematics, and thus no veto mask is applied. The final number of galaxies for each redshift bin after implementing the veto mask is shown in Table~\ref{tab:number of galaxies}. In each case, the same veto mask used for the lenses is applied to the sources, this way reducing potential geometric effects that might affect our measurements. 

\subsection{Photometric redshift distributions}\label{subsec:N(z)}

We test the robustness of the galaxy bias measurement against different \pz codes to compute the $N(z)$. For the lenses, we use BPZ to estimate the redshift distribution of the bins defined with BPZ, and analogously for TPZ, in agreement with Cr16 and G16 (top panel of Fig.~\ref{fig: nofz}). Regarding the sources, we pick all four \pz codes described in Sec.~\ref{subsec:photoz codes} to estimate the $N(z)$ (bottom panel of Fig.\ref{fig: nofz}). We choose SkyNet to be the fiducial code for this purpose, in consistency with other DES-SV analysis \citep{Becker2015, TheDarkEnergySurveyCollaboration2015}. For clarity purposes, in Fig.~\ref{fig: nofz} we show the normalized counts for the redshift bins without applying the veto mask, which is different for each bin and is described in Sec.~\ref{subsec: mask}. Nevertheless, over the analysis we use the $N(z)$'s corresponding to the redshift bins with a mask applied (the same to both the lens and source bins), although the differences with the ones shown in Fig.~\ref{fig: nofz} are unnoticeable in practice. 
\section{Measurement methodology}
\label{sec:measurement methodology}

\subsection{Measurement of the tangential shear and the cross-component}\label{subsec: tangential shear measurement}

In this section we describe how the tangential shear, the main observable of the galaxy-galaxy lensing signal, is measured. In a similar way, we can also estimate the (expected to be null) cross-component of the shear, which is a useful test of possible systematic errors in the measurement. 

First, for a given lens-source pair $j$ of galaxies we can define the tangential component of the ellipticity $\epsilon_t$ and the cross-component $\epsilon_\times$ as
\begin{linenomath}
\begin{equation} \label{eq: tangential and cross ellipticity}
\epsilon_{t, j} = -\text{Re} \left[ \epsilon_j \text{e} ^{-2i\phi_j} \right] \quad , \quad \epsilon_{\times, j}= -\text{Im} \left[ \epsilon_j \text{e} ^{-2i\phi_j} \right] ,
\end{equation}\end{linenomath} 
where $\phi_j$ is the position angle of the source galaxy with respect to the horizontal axis of the Cartesian coordinate system centered at the lens galaxy. It is convenient to consider the shear components in a  reference frame rotated with respect to the Cartesian one, because of the manner background galaxies appear distorted by the foreground mass distribution. 

This distortion is expressed in two different ways: with an anisotropic stretching turning a circle into an ellipse, i.e., a change in the shape of the galaxy quantified by the shear $\gamma$, and also with an isotropic increase or decrease of the observed size of a source image (magnification) quantified by the convergence $\kappa$. Since both effects cannot be disentangled unless magnification studies are considered, we can only measure what is called the reduced shear $g$, which includes the effect of magnification. The reduced shear is related to the shear by:
\begin{linenomath} \begin{equation}\label{eq: reduced shear}
g = \frac{\gamma}{1-\kappa}.
\end{equation}\end{linenomath} 
Since in the weak lensing regime both the shear and the convergence are much smaller than unity, $ g\simeq \gamma$. On the other hand, the observed ellipticity $\epsilon$ of a galaxy can be related to its intrinsic ellipticity $\epsilon^s$ by the following expression \citep{Seitz1997}, when the reduced shear $g$ is much smaller than one:
\begin{linenomath} \begin{equation}\label{eq: observed ellipticity}
\epsilon = \frac{\epsilon^s + g }{1+g^*\epsilon^s},
\end{equation}\end{linenomath} 
where the asterisk "*" denotes complex conjugation. Hence, the observed ellipticity can be approximated as the sum of the intrinsic ellipticity and the part due to shear: $\epsilon \simeq \epsilon^s + \gamma $. The effect this approximation might cause on the results is discussed in Sec.~\ref{subsec: reduced shear}.

Moreover, assuming intrinsic ellipticities are randomly aligned, which might not always be the case (see Sec.~\ref{subsec: intrinsic alignments}), we can obtain the shear by averaging the ellipticity over a sample of galaxies $ \gamma \simeq \left< \epsilon \right> $. In our case, we grouped the galaxy pairs in 11 log-spaced angular separation bins from 4 to 100 arcminutes. Thus, including the weighting factors from Eq.~(\ref{eq: weights}) the tangential shear and cross-component are measured using \texttt{TreeCorr}\footnote{\texttt{https://github.com/rmjarvis/TreeCorr}} \citep{Jarvis2004} in the following way:
 \begin{linenomath} \begin{equation}\label{eq: gamma_t estimation}
\gamma_{\alpha} \left(\theta\right) = \frac{\sum_j \omega_j \epsilon_{\alpha, j}  }{\sum_j \omega_j},
\end{equation}\end{linenomath} 
where $\alpha$ denotes the two possible components of the shear from (\ref{eq: tangential and cross ellipticity}). However, a typical galaxy is distorted less than $1\%$ while the intrinsic typical ellipticity is of the order of $20\%$. Therefore, to measure the shear with a significant signal to noise a large number of galaxies is needed to average out the tangential component of the intrinsic ellipticity, and hence bring down the shape noise in the measurement of the tangential shear. 
  
Other possible sources of inaccuracy are shear systematics. However, if the source galaxies are distributed isotropically around the lenses, additive shear systematics should average to zero. Still, due to edge and mask effects, there is a lack of symmetry on the sources distribution around the lenses. This effect can be accounted for by removing from the main galaxy-galaxy lensing measurement the signal measured around random points, which will capture the geometric effects of additive shear systematics. Thus, our final estimator for the tangential shear is:
\begin{linenomath} \begin{equation} \label{eq: random points subtraction}
 \gamma_t (\theta) = \gamma_t (\theta) _{\text{Lens}} -  \gamma_t (\theta) _{\text{Random}}.
 \end{equation}\end{linenomath}  
Multiplicative shear bias can still be present and we assess them as explained in Sec.~\ref{subsec: multiplicative shear biases}.

\subsection{Covariance matrix}\label{subsec: covariance matrix}

The covariance matrix for the tangential shear is estimated using a combined approach between the jackknife method and a theory estimate, as in Cr16. In this section we describe both procedures and how we merge them. 

\begin{figure}
	  \includegraphics[width=0.5\textwidth]{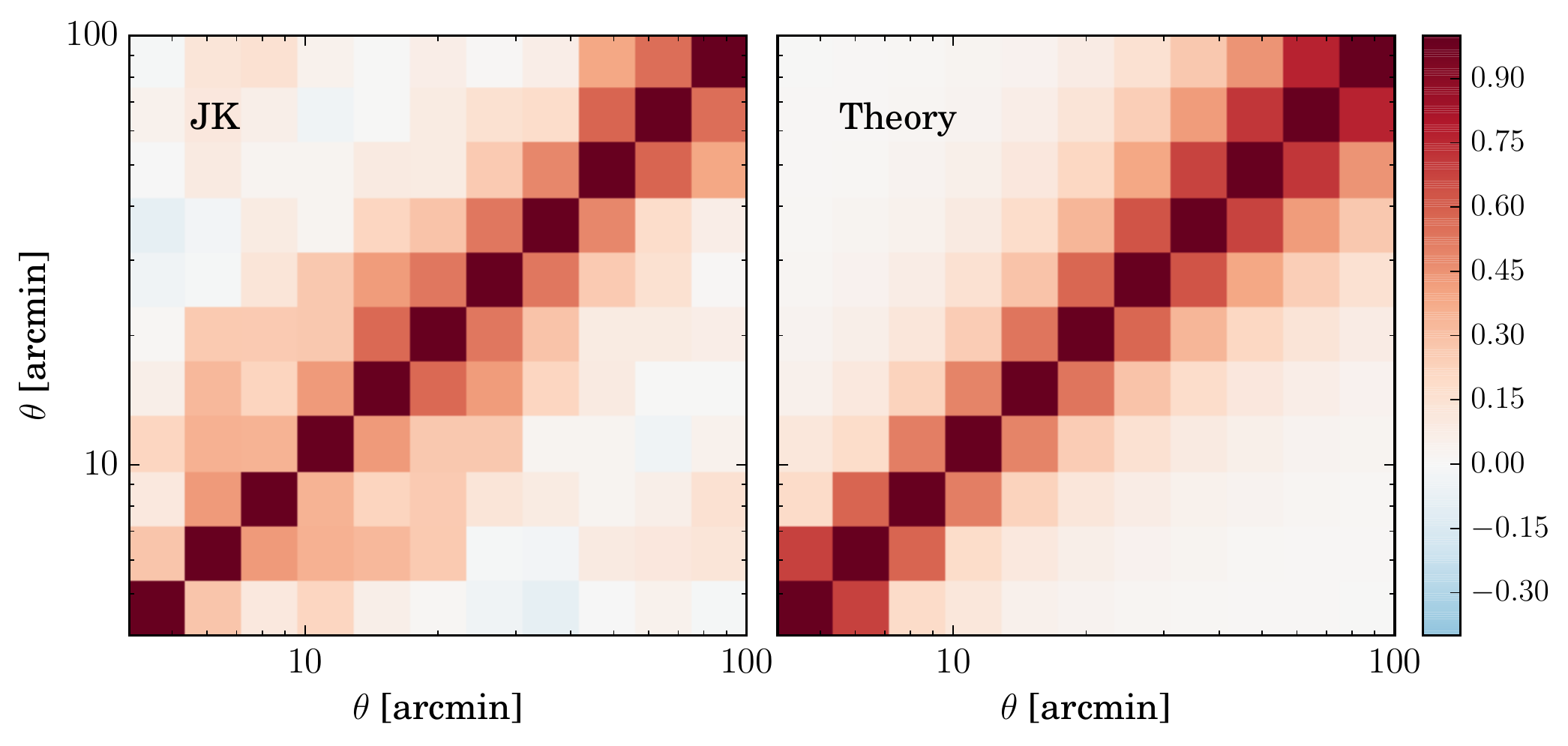} 
    \caption{Correlation matrix (normalized covariance) for the mid-$z$ lens bin defined with BPZ. \textit{Left panel:} Estimated with the jackknife method (Sec.~\ref{subsubsec: jackknife method}). \textit{Right panel:} Estimated with a theoretical modelling (Sec.~\ref{subsubsec: modelling covariance}). }\label{fig: normalized covariance}
\end{figure}

\begin{figure}
	  \includegraphics[width=0.5\textwidth]{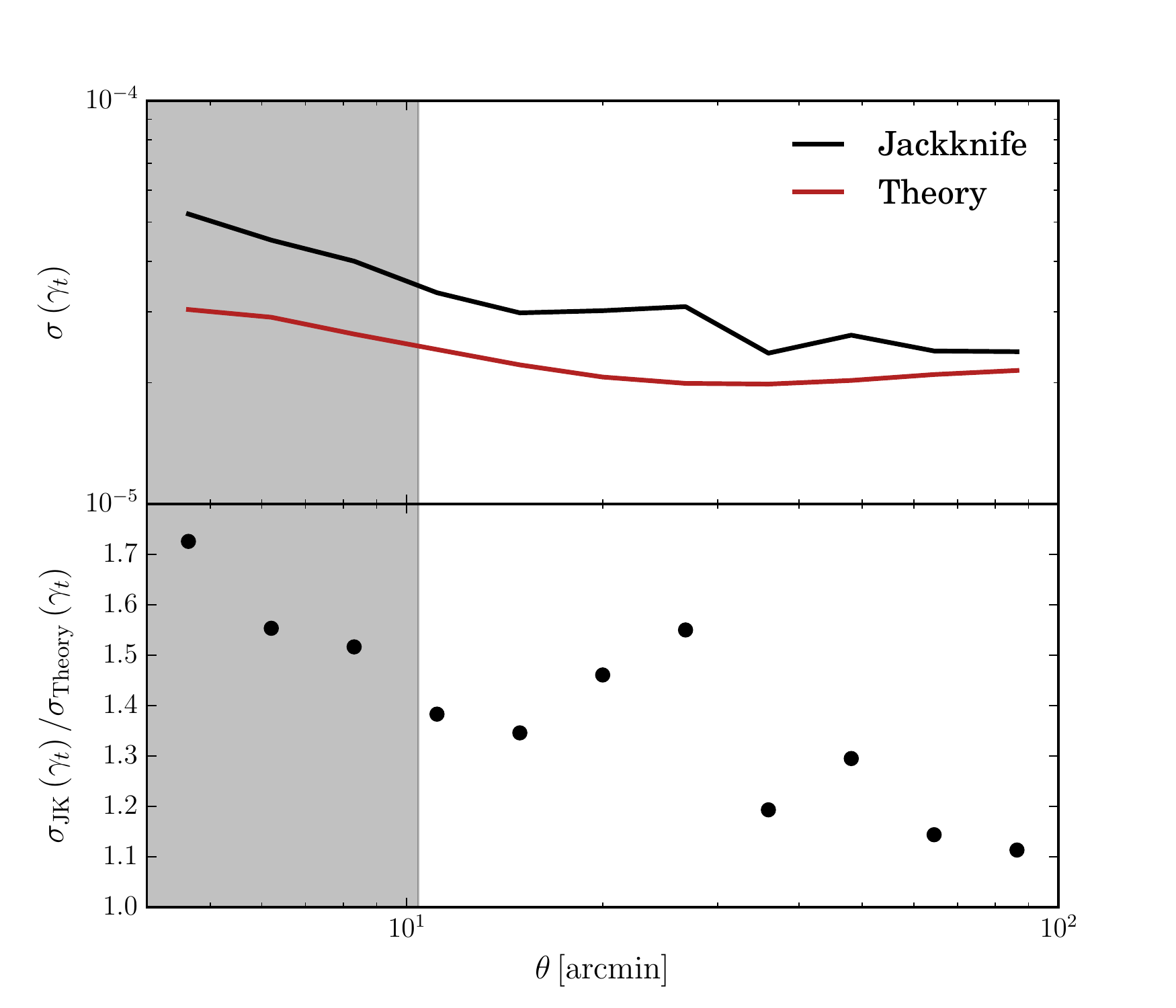} 
    \caption{Comparison of the diagonal elements of the covariance from the jackknife method and the theory predictions for the mid-$z$ lens bin defined with BPZ.}\label{fig: compare diagonal}
\end{figure}

\subsubsection{Jackknife method} \label{subsubsec: jackknife method}

The jackknife method (see for instance \citealp{Norberg2009}) is a resampling technique especially useful to estimate covariances. We divide the SPT-E area into 100 spatial jackknife regions of $\sim 1 $ sq. deg., comparable to the maximum angular scales considered, of 100 arcminutes, using the \texttt{kmeans} algorithm\footnote{\texttt{https://github.com/esheldon/kmeans\_radec}}. We tested the case with $N=50$ and obtained comparable results, with error fluctuations at the 10\% level. Then, we perform the galaxy-galaxy measurement multiple times with a different region omitted each time to make $N=100$ jackknife realizations. The covariance matrix of the tangential shear estimated with jackknife is:
\begin{linenomath} \begin{equation}
\textsf{Cov}_{ij}^{\text{JK}}\, (\gamma_{i}, \gamma_{j})= \frac{(N-1)}{N} \, \sum_{k=1}^N \, \left[ \, (\gamma_{i})^k - \overline{\gamma_{i}}\,\right] \, \left[ \, (\gamma_{j})^k - \overline{\gamma_{j}}\, \right],
\end{equation}\end{linenomath} 
where $\gamma_{i}$ represents either $\gamma_t (\theta_i)$ or $\gamma_\times (\theta_i)$ and $(\gamma_{i})^k$ denotes the measurement from the $k^{th}$ realization and the $i^{th}$ angular bin: $  \gamma_t (\theta_i)^k$. Then, the mean value is 
\begin{linenomath} \begin{equation}
\overline{\gamma_{i}}= \frac{1}{N} \, \sum_{k=1}^N \, (\gamma_i)^k.
\end{equation}\end{linenomath} 

\citet{Clampitt2016} validated the jackknife method on simulations using 50 jackknife regions on a similar patch of the sky. In there, as well as in \citet{Shirasaki2016}, it was found that the jackknife method overestimates the true covariance on large scales, where the covariance is no longer dominated by shape noise. However, recently, \citet{Singh2016} performed an extended study on galaxy-galaxy lensing covariances which concluded that subtracting the tangential shear around random points, as we do in this work, removes the overestimation of jackknife errors that was previously seen in \citet{Clampitt2016} and \citet{Shirasaki2016}. That further validates the usage of jackknife covariances in this analysis.

A jackknife-estimated normalized covariance for a particular choice of \pz and shear catalog is shown on the left panel of Fig.~\ref{fig: normalized covariance}. 

\subsubsection{Analytic covariance}\label{subsubsec: modelling covariance}

We can also model the tangential shear covariance matrix to obtain a less noisy estimate. Theoretical estimates of galaxy-galaxy lensing covariances have been studied in \citet{Jeong2009} and in  \citet{Marian2015}:
\begin{linenomath} \begin{equation}\label{eq: theory covariance}
\begin{split}
\textsf{Cov}_{ij}^{\text{TH}} \left( \gamma_{t, i}, \gamma_{t, j} \right) = \frac{1}{4\pi f_\text{sky}} \int \frac{l \text{d}l }{2\pi }	\bar{J}_2(l  \theta_i) \bar{J}_2 (l\theta_j) \\
\times \left[ \mathcal{C}^2_{g\kappa} (l) +  
\left( \mathcal{C}_{gg}(l) + \frac{1}{n_{\text{L}} }\right)
\left( \mathcal{C}_{\kappa \kappa}(l) + \frac{\sigma^2_\text{SN}}{n_{\text{S}} }
\right)\right],
\end{split}
\end{equation}\end{linenomath} 
where $f_\text{sky}$ is the fraction of the sky covered; $n_\text{L}$ and $n_\text{S}$ are the effective number density of the lenses and the sources, respectively, defined in \citet{Jarvis2015}; $\mathcal{C}_{g \kappa}(l)$, $\mathcal{C}_{gg}(l)$ and  $\mathcal{C}_{\kappa \kappa}(l)$ are the line of sight projections of the galaxy-matter, galaxy-galaxy and matter-matter power spectrum, respectively, obtained using \texttt{Halofit} \citep{Smith2003, Takahashi2012} with \texttt{CosmoSIS}; $\sigma_\text{SN}$ is the shape noise per component and $\bar{J}_2$ are the bin-averaged Bessel functions of order two of the first kind, defined as:
\begin{linenomath} \begin{equation}\label{eq: J2}
 \bar{J}_2(l\theta_i)  \equiv \frac{2\pi}{A(\theta_i)}\int _{\theta_{i, \text{min}}}^{\theta_{i, \text{max}}} J_2(l\theta) \theta \text{d} \theta,
\end{equation}\end{linenomath}    
where $A(\theta_i) = \pi \left( \theta^2_{i, \text{max}} - \theta^2_{i, \text{min}}\right) $ is the area of the bin annulus. We integrate (\ref{eq: theory covariance}) over $1 \leq l < 4000$, which covers the range of scales used in this work. A theory estimated normalized covariance matrix is shown on the right panel of Fig.~\ref{fig: normalized covariance}. It is much smoother than the jackknife estimation (left panel), particularly far from the diagonal. 

\subsubsection{Combined approach}

As shown on the left panel of Fig.~\ref{fig: normalized covariance}, the jackknife method gives a rather noisy estimate of the off-diagonal elements, due to the impossibility of increasing the number of realizations without being forced to use excessively small jackknife regions. However, it is relevant to obtain good estimates of the off-diagonal terms since adjacent angular bins are highly correlated. Moreover, the inverse of a noisy, unbiased covariance is not an unbiased estimator of the inverse covariance matrix \citep{Hartlap2007}, which is needed to fit the galaxy bias (see Sec.~\ref{subsec: galaxy bias fit}). We improve the estimation of the covariance by obtaining a smooth correlation matrix from theory estimation, shown on the right panel of Fig.~\ref{fig: normalized covariance}.  

On the other hand, concerning now the overall normalization of the covariance matrix, the jackknife procedure is capable of capturing effects that potentially exist in the data and cannot be derived from theory, such as shear systematics or mask effects, and can also reproduce non-linearities although we expect these to be small over the scales used. Indeed, in G16, the jackknife method was found to perform better on the diagonal elements over the theory estimates when compared to a covariance matrix derived from an N-body simulation. Also, the diagonal elements from jackknife are in principle better estimated than the off-diagonal ones, since there is more signal-to-noise in the diagonal. Then, following Cr16, we choose to combine both methods by normalizing the theory-estimated covariance with the diagonal elements of the jackknife covariance:
\begin{linenomath} \begin{equation}
\textsf{Cov}^\text{COMB}_{\theta_i, \theta_j} = \textsf{Corr}^\text{TH}_{\theta_i, \theta_j} \sigma^\text{JK} _{\theta_i} \sigma^\text{JK}_{\theta_j}.
\end{equation}\end{linenomath} 
The comparison between the diagonal elements from the jackknife method and from theory predictions can be found in Fig.~\ref{fig: compare diagonal}. The jackknife procedure yields larger diagonal elements for the covariance as a result of including additional sources of uncertainties as discussed above.

\subsection{Galaxy bias fit}\label{subsec: galaxy bias fit}

We can now put together all the required ingredients to measure the product of the galaxy bias $b$ times the cross-correlation coefficient $r$: the measured tangential shear (Sec.~\ref{subsec: tangential shear measurement}), the modelled tangential shear (Sec.~\ref{sec:method}) and the covariance matrix (Sec.~\ref{subsec: covariance matrix}). Then, $b\cdot r$ is measured minimizing the following $\chi^2$:
\begin{linenomath} \begin{equation}\label{eq: galaxy bias fit}
\begin{split}
\chi^2\, (b\cdot r) &= \sum_{\theta, \theta'} \left( \gamma_t (\theta) - b\cdot r \  
\gamma_t^\text{TH} (\theta) \right)\textsf{Cov}^{-1}\left(\theta, \theta'\right) \\
& \times  \left(\gamma_t (\theta') - b\cdot r \ 
\gamma_t^\text{TH} (\theta') \right),
\end{split}
\end{equation}\end{linenomath} 
where $\gamma_t^{\text{TH}}$ assumes $b\cdot r = 1$.

\begin{figure*}
    \includegraphics[width=\textwidth]{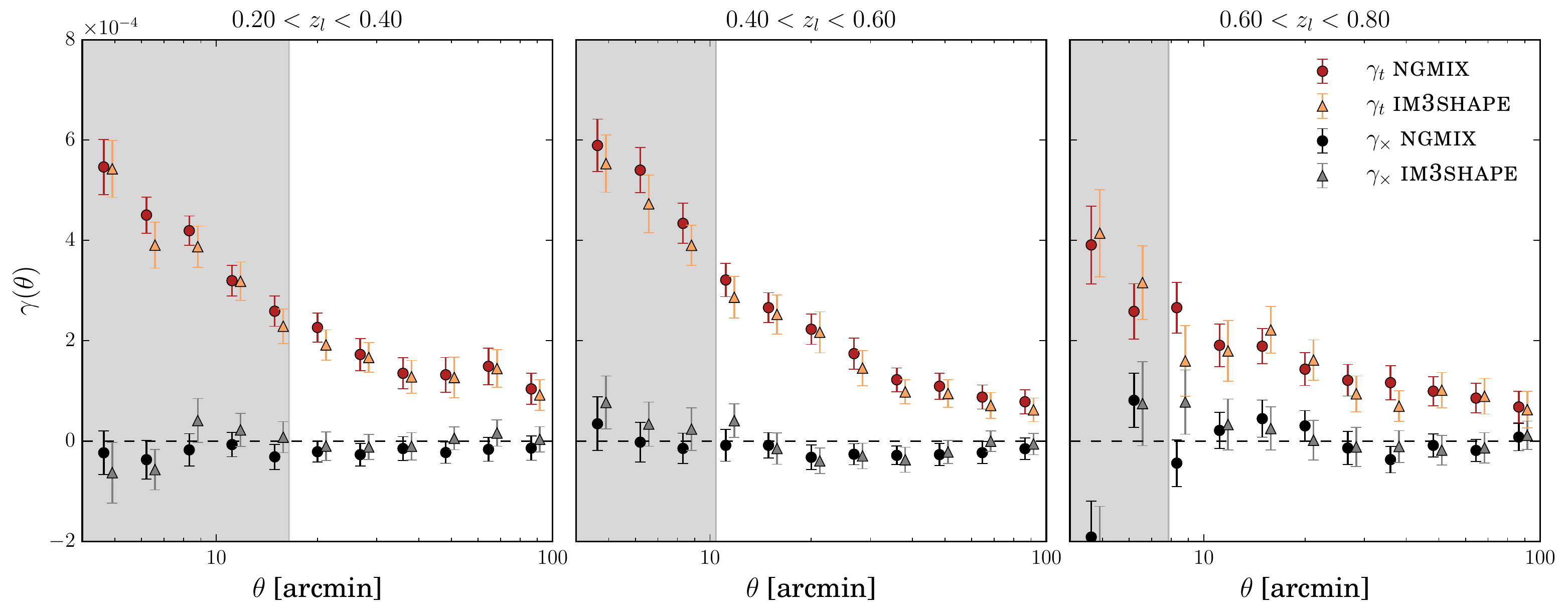}
    \caption{Tangential shear $\gamma_t$ as a function of angular scales from \ngmix (red points) and from \im (orange triangles) for the three lens redshift bins defined with BPZ. Note that the measurements from \ngmix and \im cannot be directly compared because of the lensing efficiency being  different in each case (see Sec.~\ref{subsec: tangential shear measurements}). Moreover, we show the cross-component (see Sec.~\ref{subsec: cross-component and random} for discussion) for both \ngmix (black points) and \im (grey triangles), which is consistent with zero in all redshift bins. The null $\chi^2$ for the cross-component are shown in Table \ref{tab:chi2 gammat gammax br}. The shaded angular scales are not considered for the final galaxy bias measurements, which are performed over the range of scales from 4 Mpc/$h$ to 100 arcmin. }\label{fig: ngmix_im3shape_gammat_gammax}
\end{figure*}

\begin{figure*}
    \includegraphics[width=\textwidth]{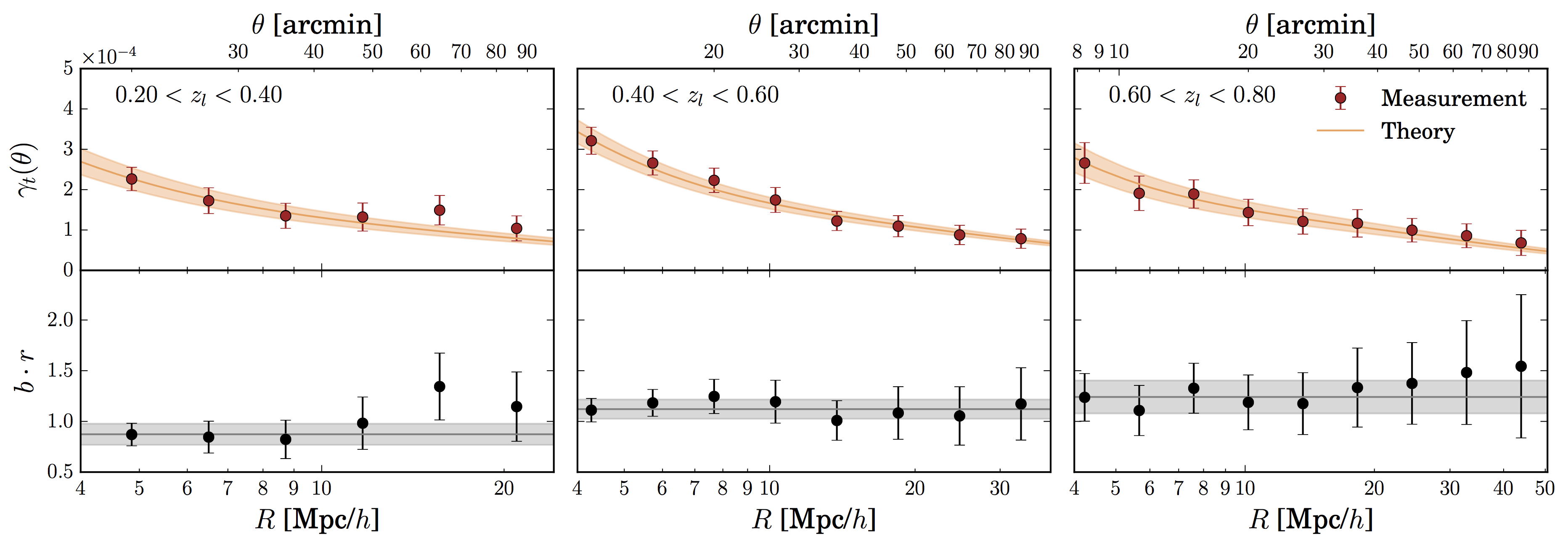}
    \caption{\textit{Top panels:} Tangential shear (red points) as a function of the transverse comoving distance $R$ (bottom axis) or the angular separation $\theta$ (top axis), together with the best-fitting theory prediction of the tangential shear modelled with \texttt{Halofit} (solid orange line) with its corresponding uncertainty band. \textit{Bottom panels:} $b\cdot r$ as a function of scale (black points) with the best-fit (black solid line) using Eq. (\ref{eq: galaxy bias fit}),  with its uncertainty band. The $N(z)$ of the source galaxies is estimated with SkyNet and the source catalog is \textsc{ngmix}, which are the fiducial choices.}\label{fig: br fits}
\end{figure*}

\subsubsection{Range of selected scales}\label{subsec: Range of selected scales}

In this section we discuss the range of scales suitable to perform the fit described above. There are some limitations that we need to consider both at large and small scales. For instance, on small scales, effects such as stochasticity, non-local bias and scale dependence take place. Since this behaviour is hard to model, we need to identify the range of scales over which the product of the galaxy bias $b$ times the correlation parameter $r$ is scale independent, free of stochasticity and non-linear effects. 

Nonetheless, it is difficult to remove all small scale information in real space, since the tangential shear at each angular scale is an integration of all multipole moments $l$ -- see Eq. (\ref{eq: gamma_t theory}) --, that is to say, it contains knowledge of all scales, weighted according to the Bessel function. Then, applying a sharp cut-off in real space does not fully erase the effects present below that cut-off. Even though there exists an alternative estimator of the tangential shear (the annular differential surface density estimator) proposed by \citet{Baldauf2010} and \citet{Mandelbaum2010} to remove all small scale information, a conservative minimum scale cut-off should be sufficient to remove enough of it for our purposes, given the current uncertainties in the measurement of the tangential shear. However, this issue might have to be addressed more carefully in future work involving larger area, which will significantly reduce the present uncertainties. 

As a result of the non-linear effects at small scales, one does not expect a constant value for either $b$ or $r$ over this range. On the other hand, in the linear bias regime, both the galaxy bias and the cross-correlation parameter can be approximated to a constant. The transition scale from the non-linear to the linear regime should be expressed as a comoving distance $R$, as opposed to an angular scale $\theta$. For galaxies acting as lenses at different redshift, the same angle $\theta$ will correspond to different distances $R$. Thus, it is convenient to convert the angle $\theta$ into the \textit{transverse comoving distance} $R$ that this angle represents. Then, for small angles, $R =  \chi_\text{lens} \, \theta,$ where $\chi_\text{lens}$ is the \textit{radial comoving distance} to the lens galaxy, which can be related to the mean redshift of the lens redshift bin. To compute $\chi_\text{lens}$ we assume the cosmology described in Sec.~\ref{sec:method}. In this manner, in Fig.~\ref{fig: br fits}, both scales are displayed: the angular separation $\theta$ at the top and the transverse comoving distance $R$ at the bottom. 

In Cr16, it was studied the scale of linear growth, which is the minimum scale where the linear and the non-linear (with \texttt{Halofit}) matter power spectrum are the same. The linear growth scale for clustering was found to be $\sim 4$~Mpc/$h$, which they adopted as a minimum comoving distance for their results.  \citet{Jullo2012} studied the scale of linear bias in the COSMOS field and determined $R = 2.3 \pm 1.5$ Mpc/$h$ to be the scale beyond which bias evolves linearly. 

On the other hand, the cross-correlation coefficient dependence on scale has been investigated in recent studies such as \citet{Baldauf2010} and \citet{Mandelbaum2013}, finding $r\simeq 1$ and scale independent on scales larger than a few virial radius of galaxy halos. \citet{Jullo2012} obtained $r$ compatible with one for $0.2 <R<15$ Mpc/$h$ and $0.2<z<1$. \citet{Hoekstra2002} measured the linear bias and cross-correlation coefficient on scales between $R = 0.2$ and 9.3 Mpc/$h_{50}$ at $z = 0.35$. They found strong evidence that both $b$ and $r$ change with scale, with a minimum value of $r\sim 0.57$ at 1 Mpc/$h_{50}$. However, on scales larger than 4 Mpc/$h_{50}$ they obtained $r$ is consistent with a value of one. In App.~A of Cr16 the cross-correlation coefficient $r$ was measured in the \texttt{MICECATv2.0} simulation, which is an updated version of \texttt{MICECATv1.0} \citep{Fosalba2015a, Fosalba2015b, Crocce2015a, Carretero2014}, including lower mass halos and thus more similar to the benchmark sample. They found the cross-correlation coefficient to be in the range $0.98 \leq r\leq 1$ for $z>0.3$ in the range of scales $12 < \theta < 120$ arcmin, which approximately correspond to a comoving minimum scale of 3 Mpc/$h$. Even though it is worth noting that different definitions of $r$ may yield different estimates and scale dependencies, it is relevant to this work the fact that, for the range of scales we use, various studies agree on a value of $r$ close to unity and showing little scale dependence.

Overall, following Cr16, we choose a conservative minimum scale cut-off of 4~Mpc/$h$. However, because the redshift bins have some non-negligible width, a significant fraction of lenses will be below the mean redshift of the bin. Thus, when converting from angular to physical scale, we are effectively including some galaxy pairs that are separated by less than 4~Mpc/$h$. We tested how important this effect is by, instead of using the mean redshift of the bin to convert from angular to physical scale, using the mean value minus one standard deviation of the redshift distribution. The variations induced by that change in the final galaxy bias measurements are at the level of 0.5-3\%, thus much lower than the statistical errors. 

Some limitations are present on large scales as well. The maximum valid scale is restricted by the size of the SV $\text{SPT-E}$ patch, of 116.2~sq.~deg. On the other hand, we are also limited by the size of the jackknife regions used in this work to estimate covariances. Then, we follow the approach used in \citet{Kwan2016} of using 100 jackknife regions and a maximum scale cut-off of 100 arcmin. 

\subsection{Non-linear bias model} \label{subsec: Nonlinear bias model}
As a further check, we have tested whether the assumption of linear bias is valid over the scales used -- larger than 4 Mpc/$h$ -- by studying the robustness of the results when using a non-linear bias scheme. In order to do so, we choose the  non-linear bias model adopted in \citet{Kwan2016} and originally developed in \citet{McDonald2006}, which is a reparametrization of the  model described in \citet{Fry1993}. In this model, the galaxy overdensity, $\delta_g$, is written as:
\begin{linenomath}
\begin{equation}
\delta_g = \epsilon + b_1 \,\delta_m + b_2\, \delta_m^2 + ...\, , 
\end{equation}
\end{linenomath}
where $\epsilon$ is the shot noise, $b_1$ is the linear bias and $b_2$ is the non-linear bias from the second order term. Then, the relationship between the galaxy-matter power spectrum and the matter power spectrum is given by
\begin{linenomath}
\begin{equation}
P_{g\delta} = b_1 \, P_{\delta \delta} + b_2 \, A(k),
\end{equation}
\end{linenomath}
since by definition $\epsilon$ is not correlated with $\delta_m$, and where $A(k)$, defined in \citet{Kwan2016}, can be calculated using standard perturbation theory. Comparing this relation to (\ref{eq: P_delta g = br P_delta delta}), we identify $b_1$ as $b\cdot r$ in the case of linear bias, corresponding to $b_2 = 0$.

Applying this non-linear bias model to the fiducial case for the three redshift bins defined with BPZ, we find that $b_1$ is compatible with the results coming from the linear bias model, and that $b_2$ is compatible with zero, for all redshift bins. The uncertainties we obtain on both $b_1$ and $b_2$ are large -- in the case of $b_1$ between 30\% and up to twice as large as for the fiducial case, depending on the redshift bin; thus, this indicates that we are lacking statistical power on these large scales to obtain competitive constrains when we introduce another parameter in the modelling. Overall, the linear bias assumption holds for scales larger than 4 Mpc/$h$, given the current uncertainties. Hence, all the results presented in the following sections are obtained using the linear bias model. 
\section{Results}
\label{sec:results}

\begin{table*}
\centering
\begin{tabular}{ lccccc}\toprule 

& Redshift bin & $\chi_{\text{null}}^2/\text{ndf} \ \left(\gamma_t \right)  $ & $\chi_{\text{null}}^2/\text{ndf} \ \left(\gamma_{\times} \right)$
 & $b\cdot r$  &$\chi_{\, \text{fit}}^2/\text{ndf} $ \\
\midrule 

\multirow{3}{*}{BPZ \textsc{ngmix} (F)} 
& $0.2\leq z_l < 0.4 $ &  72.2/6  &  3.6/6  & $0.87\pm 0.11$   & 3.4/5 \\ 
& $0.4\leq z_l < 0.6 $ &  138.2/8  &  4.9/8  &  $1.12 \pm 0.16$  & 2.2/7 \\ 
&$0.6\leq z_l < 0.8 $ &  59.5/9  &  8.7/9  &  $1.24 \pm 0.23$  & 1.4/8\\ 
[0.2cm] 

\multirow{3}{*}{BPZ \textsc{im3shape}} 
& $0.2\leq z_l < 0.4 $ &  56.8/6  &  0.95/6  & $0.79 \pm 0.12$   & 3.8/5 \\ 
& $0.4\leq z_l < 0.6 $ &  79.0/8  &  5.2/8  &  $1.03 \pm 0.17$  & 3.3/7 \\ 
&$0.6\leq z_l < 0.8 $  &  35.1/9  &  3.0/9  &  $1.08 \pm 0.25$  & 7.5/8\\ 
[0.2cm] 

\multirow{3}{*}{TPZ \textsc{ngmix} (F)} 
& $0.2\leq z_l < 0.4 $ & 61.1/6  &  2.3/6  & $0.77 \pm 0.11$ & 1.9/5\\ 
& $0.4\leq z_l < 0.6 $ & 124.5/8 &  4.6/8 & $1.40 \pm 0.21$ & 1.9/7\\ 
& $0.6\leq z_l < 0.8 $ &  93.0/9  & 4.1/9 & $1.57 \pm 0.27$ & 0.82/8 \\ 
[0.2cm] 

\multirow{3}{*}{TPZ \textsc{im3shape}} 

& $0.2\leq z_l < 0.4 $ &  48.8/6  &  0.98/6  &  $0.78 \pm 0.13$  & 4.5/5 \\ 
& $0.4\leq z_l < 0.6 $ &  93.4/8  &  5.9/8  &  $1.34 \pm 0.22$  & 0.83/7 \\ 
& $0.6\leq z_l < 0.8 $ &  53.5/9 &  4.5/9  &  $1.36 \pm 0.28$  & 6.7/8 \\ 
\bottomrule
\end{tabular}
\caption{ \label{tab:chi2 gammat gammax br} Best-fitting galaxy bias results ($b\cdot r$) for the four main different combinations of \pz codes and shear catalogs, shown also in Fig.~\ref{fig: comparison br results}. For instance, BPZ \textsc{ngmix} stands for lens redshift bins defined with BPZ and \ngmix as the source catalog. $\chi_{\text{null}}^2/\text{ndf} \ \left(\gamma_t \right)$ is the null $\chi^2$ of the tangential shear over the number of degrees of freedom, covering the range of scales from 4 Mpc/$h$ to 100 arcmin (not shadowed region in Fig.~\ref{fig: ngmix_im3shape_gammat_gammax}), and the same for the cross-component $\gamma_\times$. $\chi_{\, \text{fit}}^2$ corresponds to the galaxy bias fit described in Sec.~\ref{subsec: galaxy bias fit}. All combinations use SkyNet as the \pz code for the $N(z)$ of the sources.}
\end{table*}

In this section we present the main results of this work. First, we introduce the tangential shear measurements to later proceed describing the galaxy bias results. 

\subsection{Tangential shear measurements}\label{subsec: tangential shear measurements}
In this subsection we discuss the measured tangential shear as a function of the angular separation, shown in Fig.~\ref{fig: ngmix_im3shape_gammat_gammax}. In that figure, BPZ is used to define the lens bins and to estimate the $N(z)$ of the lenses, and SkyNet is used to estimate the $N(z)$ of the sources. Also, the shaded angular scales from Fig.~\ref{fig: ngmix_im3shape_gammat_gammax} 	are not considered for the final galaxy bias measurements, which are performed on the range of separations from 4~Mpc/$h$ to 100~arcmin (see Sec.~\ref{subsec: Range of selected scales}).

We measured the tangential shear using two different shear catalogs: \ngmix and \im (see Sec.~\ref{subsec: The sources: shear catalogs}), which correspond to different galaxy samples, and thus, different redshift distributions. Then, the measurements for the tangential shear for \ngmix and \im cannot be directly compared because of the lensing efficiency being different in both cases. Nevertheless, we will be able to compare galaxy bias measurements (see Fig.~\ref{fig: comparison br results}), which are independent of the source sample redshift distribution, assuming an unbiased estimation of the $N(z)$ (see Sec.~\ref{subsec: photoz errors}). We choose \ngmix to be the fiducial shear catalog as it includes more galaxies, inducing less shape noise in the measurement. This effect is especially noticeable at small scales, where the shape noise contribution dominates the error budget. 

In Table~\ref{tab:chi2 gammat gammax br} we display the $\chi^2$ of the null hypothesis and the number of degrees of freedom for the tangential shear signal over the selected range of scales, for all the different redshift bins and \pz codes. We measure a non-zero tangential shear signal over the aforementioned scales for all different \pz and shear catalog choices. Calculating the signal-to-noise as $\text{S/N}= \sqrt{\chi^2_\text{null}(\gamma_t) - N_\text{bin}}$, where $N_\text{bin}$ is the number of angular bins considered for the galaxy bias measurements, the maximum S/N we obtain is 11.4 for the mid-$z$ BPZ + \ngmix bin, and the minimum is 5.1 for the high-$z$ BPZ + \im bin.

\subsection{Galaxy bias results}\label{subsec: galaxy bias results}

In Fig.~\ref{fig: br fits} we present the galaxy bias fits for the fiducial \pz codes and shear catalog, using BPZ to define the lens bins and SkyNet to estimate the $N(z)$ of the sources for the three redshift bins. On the top panels, we show the measured tangential shear together with the best-fitting theory prediction for the tangential shear over the scales of interest, which are from 4~Mpc/$h$ to 100~arcmin. We display the comoving distance $R$ on the bottom axis and the angular scale $\theta$ on the top axis. On the bottom panels, we show the galaxy bias as a function of separation with the best-fitting value, obtained using Eq. (\ref{eq: galaxy bias fit}). 

\begin{figure}
    \centering
      \hspace*{-5mm}
        \includegraphics[width=0.5\textwidth]{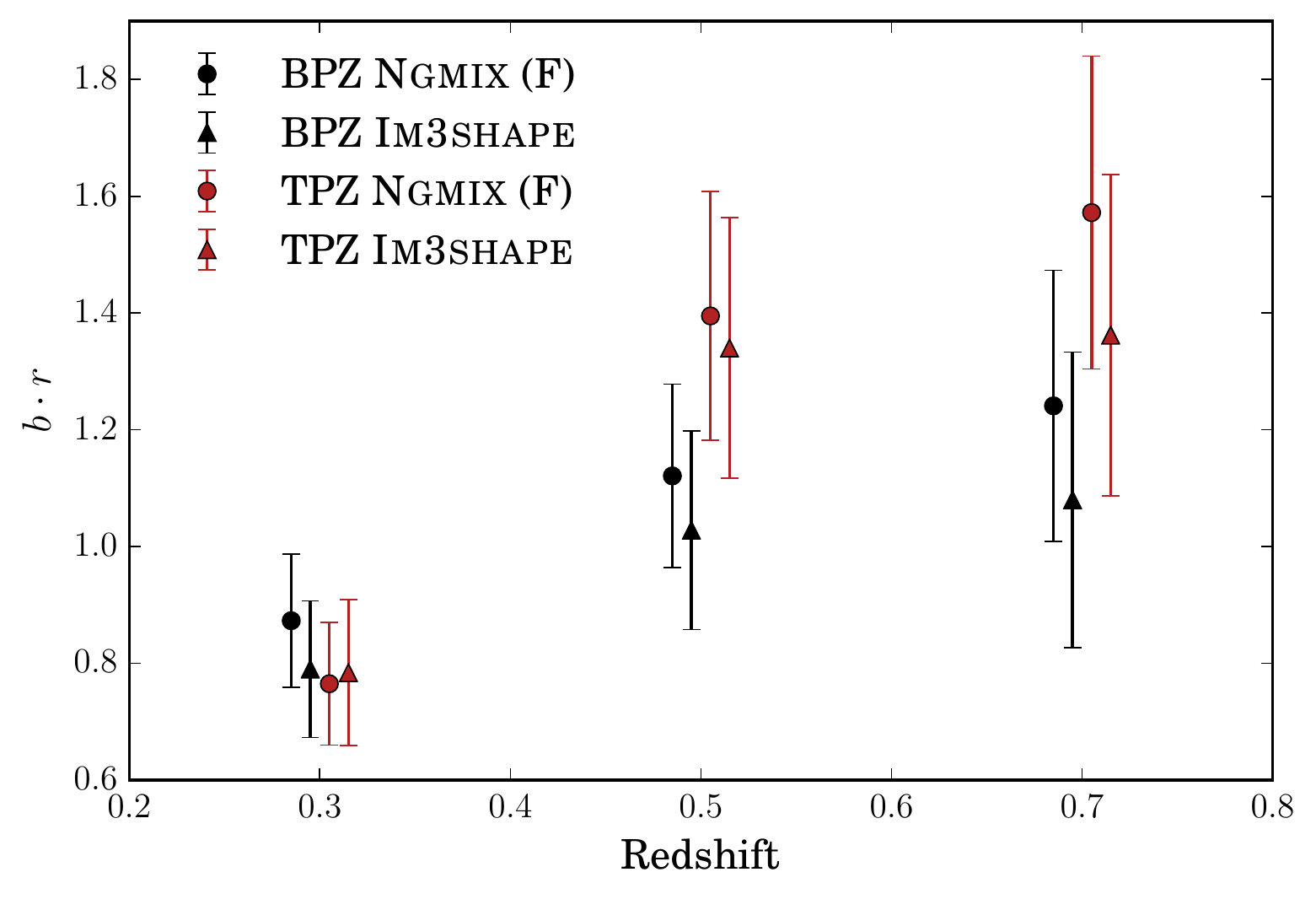}
    \caption{Fiducial galaxy bias results (F) along with a comparison between results obtained with different combinations of \pz codes and shear catalogs. For instance, BPZ \ngmix stands for lens redshift bins defined with BPZ and \ngmix as the source catalog. All combinations use SkyNet as the \pz code for the $N(z)$ of the sources. The points have been offset horizontally for clarity purposes. }\label{fig: comparison br results}
\end{figure}

In Figs.~\ref{fig: comparison br results} and \ref{fig: br results source photoz} we show our fiducial galaxy bias results (BPZ + SkyNet with \textsc{ngmix} and TPZ + SkyNet with \textsc{ngmix}), along with the rest of combinations of \pz codes and shear catalogs. The results from Fig.~\ref{fig: comparison br results} are also presented in Table~\ref{tab:chi2 gammat gammax br}, together with the best-fit $\chi^2$. The low $\chi^2$ values in some of the cases might be due to an overestimation of the uncertainties given by the jackknife method, which will only lead to more conservative conclusions. In Fig.~\ref{fig: comparison br results} we compare the results varying the \pz code for the $N(z)$ of the lenses. However, this comparison is not straightforward due to the fact that galaxies in each lens sample are not the same -- they have been defined using either BPZ or TPZ to directly compare with the measurements from Cr16 and G16, which use the same binning. Actually, the number of common galaxies in each redshift bin is: 273133, 406858, 348376, compared to the number of galaxies in each bin, given in Table~\ref{tab:number of galaxies}. Hence, the galaxy bias might actually \textit{be} different for each case. Nevertheless, even though we observe variations in the galaxy bias values, these differences are within the uncertainties. In Fig.~\ref{fig: comparison br results} we also compare the galaxy bias results from using the two different shear catalogs \textsc{ngmix} (fiducial) and \textsc{im3shape}. We obtain agreement between the results from the two shear samples.  

In Fig.~\ref{fig: br results source photoz} we test the robustness of our results under the choice of the \pz code used to estimate the $N(z)$ of the sources. We detect variations in the galaxy bias up to 6\%, 9\% and 14\% for the three redshift bins, respectively. We include this source of systematic in the error budget as described in Sec.~\ref{subsec: photoz errors}.

\subsubsection{Galaxy bias evolution}

In Figs.~\ref{fig: comparison br results} and \ref{fig: br results source photoz}, we observe the evolution of the galaxy bias with redshift, in all combinations of \pz codes and shear catalogs. There are two main reasons for this evolution. 

First, at high redshift ($z\sim 3$), galaxies form at special locations in the density field where they already trace the network of filaments emerging in the dark matter distribution. The dark matter correlation function grows in time as mass moves into this network from the surrounding regions, but the structure traced by galaxies stays relatively unchanged, and the galaxy correlation function is only weakly dependent on redshift (e.g. \citealt{Weinberg2004a}). Then, because the dark matter correlation function does evolve in time, we expect the galaxy bias to evolve as well. More precisely, we expect the galaxy bias to be larger than one at high redshift, which means that the galaxy distribution is more clustered than the dark matter distribution. 

Secondly, since we are studying a magnitude-limited sample of galaxies, in average we are naturally observing a higher luminosity sample at higher redshift. We find an increase of slightly more than a unit in absolute magnitude in the $i$ band (corresponding to a factor of $\sim$3 in luminosity) between the low-$z$ and the high-$z$ lens bins. Since more luminous galaxies tend to be more biased, we would already expect the bias to increase with redshift even without intrinsic bias evolution.

\section{Systematic effects in galaxy-galaxy lensing}
\label{sec:systematics}

In this section we explore the different systematic effects that can potentially plague our galaxy-galaxy lensing measurements. For that purpose, we perform some null tests on the data and present a series of calculations, some of them using previous analyses on the same data sample, which are all described in detail next. A summary of the significant contributions from these systematics to the total error budget is presented in Table~\ref{tab:systematics}.

\subsection{Cross-component and tangential shear around random points}\label{subsec: cross-component and random}

The cross-component of the shear $\gamma_{\times}$, which is rotated 45 degrees with respect to the tangential shear, should be compatible with zero if the shear is only produced by gravitational lensing, since the tangential shear captures all the galaxy-galaxy lensing signal. Hence, measuring $\gamma_\times$ provides a test of systematic errors, such as point-spread function (PSF) related errors, which can leak both into the tangential and cross-components of the galaxy shear. PSF leakage could arise from errors in the PSF model, as well as residual errors in correcting the PSF ellipticity to estimate the galaxy shear; such correction is done by analyzing the shape of stars in the field. In Sec.~\ref{subsec: tangential shear measurement} we describe how the measurement of the cross-component of the shear is performed and in Fig.~\ref{fig: ngmix_im3shape_gammat_gammax} is shown for the foreground redshift bins defined with BPZ. 
In order to test whether the cross-component of the shear is compatible with zero, we compute the null $\chi^2$ statistic:
\begin{linenomath}
\begin{equation}
\chi^2_\text{null} = \boldsymbol{\gamma}_\times^T \, \boldsymbol{\cdot} \, \textsf{Cov}^{-1} \boldsymbol{\cdot} \, \boldsymbol{\gamma}_\times,
\end{equation}
\end{linenomath}
where the covariance matrix for the cross-component is estimated with the jackknife method, described in Sec.~\ref{subsubsec: jackknife method}. Since jackknife covariance matrices comprise a non-negligible level of noise, in order to obtain an unbiased estimate of the inverse covariance a correction factor of $(N - p -2)/(N-1)$ has to be applied to the inverse covariance, where $N$ is the number of JK regions and $p$ is the number of angular bins \citep{Hartlap2007, Kaufman1967}.
This factor corrects for the fact that, as mentioned in Sec. 4.2.3, the inverse of an unbiased but noisy estimate of the covariance matrix is not an unbiased estimator of the inverse of the covariance matrix. In Table~\ref{tab:chi2 gammat gammax br} we show all the $\chi^2_\text{null}$ values for each of the redshift bins, which are all consistent with zero. 

A second test for galaxy-galaxy lensing is the measurement of the tangential shear around random points. This measurement tests the importance and possible contribution from geometrical effects in the signal. Although our estimator of galaxy-galaxy lensing in Eq. (\ref{eq: random points subtraction}) includes the subtraction of the random points signal, it is useful to check that this correction is small. This measurement was presented in \citet{Clampitt2016} with the same sources that we use, and they found the signal to be consistent with zero.

\subsection{Photo-\boldmath{$z$} errors} \label{subsec: photoz errors}

In this section, we discuss the impact of \pz errors on $b\cdot r$ uncertainties. Particularly, we focus on the effect caused by an overall shift on the redshift distribution of the sources. We approach this subject by following the recommendation from \citet{Bonnett2015} of adopting a Gaussian prior of width 0.05 for the shift $\delta_i$ on the mean of the distribution of the source galaxies: $N_i(z)\rightarrow N_i(z-\delta_i) $. We draw 1000 realizations of the $\delta_i$, measuring the galaxy bias each time. Then, we add the standard deviation of the galaxy bias values in quadrature to the statistical error budget. 

\begin{figure}
      \hspace*{-5mm}
        \includegraphics[width=0.5\textwidth]{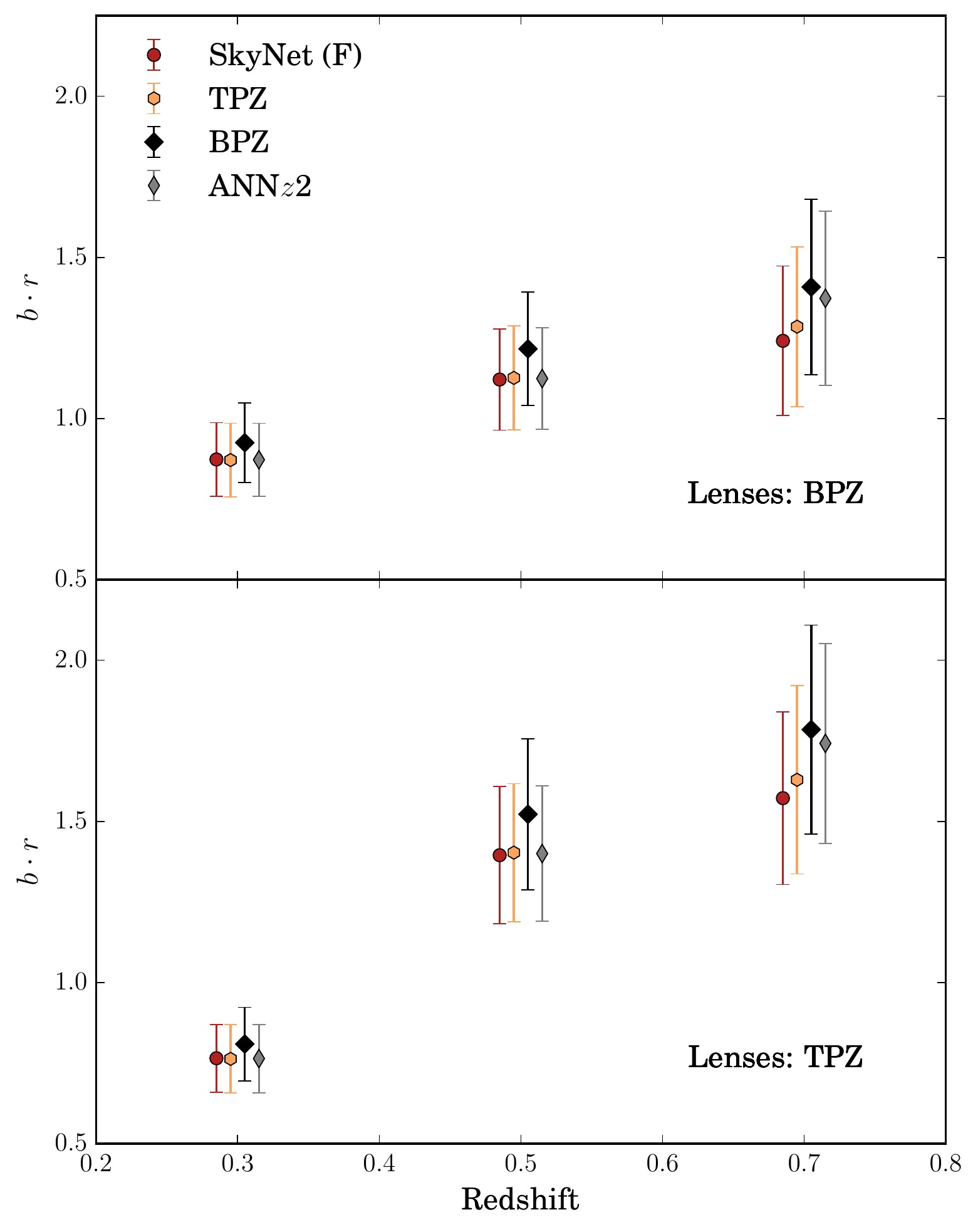}
    \caption{Galaxy bias results varying the \pz code to estimate the redshift distribution of the source sample. For instance, for the red points, the $N(z)$ of the sources is estimated with SkyNet (Fiducial). We observe good agreement among the different results. \textit{Top panel:} Lens redshift bins are defined with BPZ and the lens $N(z)$ is estimated using BPZ as well. \textit{Bottom panel:} The same with TPZ. \ngmix is the source catalog used for these results. The points have been offset horizontally for clarity purposes.} 
        
\label{fig: br results source photoz}
\end{figure}

Including the \pz error contribution represents a fractional increase of 7\%, 38\% and 42\% to the galaxy bias statistical uncertainty for each redshift bin from low to high redshift (see Table~\ref{tab:redshift bins}), after averaging over all different \pz choices (for \textsc{ngmix} only).  Although almost the same source redshift distributions are used for the first ($0.2\leq z_l<0.4$ and $0.55<z_s<1.3$) and second redshift bin ($0.4\leq z_l<0.6$ and $0.55<z_s<1.3$) -- not exactly the same because the veto masks are applied, which are different for each bin -- the increase of the errors is significantly larger for the second redshift bin because of the geometrical factors involved. On the other hand, on average these \pz uncertainties represent a 5\%, 8\% and 12\% of the galaxy bias measured in each bin, similar to the 6\%, 9\% and 14\% of maximum variation of the galaxy bias, with respect to the fiducial value, when changing the \pz code to estimate the $N(z)$ of the source galaxies (see Sec.~\ref{sec:results} and Fig.~\ref{fig: br results source photoz}). 

\subsection{Reduced shear and magnification} \label{subsec: reduced shear}

\begin{table}
\centering
\begin{tabular}{ccccc}\toprule 
 Lens $z$-bin & $\sigma_\text{stat}/(b\cdot r)$ & $\sigma_\text{p-$z$}/(b\cdot r)$ &  $\sigma_\text{m}/(b\cdot r)$ & $\sigma_\text{IA}/(b\cdot r)$   \\
\midrule 
$0.2 \leq z_l < 0.4$ & 12\%$^*$  & 5\%$^*$    & 2\%   & $-$\\
$0.4 \leq z_l < 0.6$ & 9\%$^*$   & 8\%$^*$    & 2\%   & 8\%\\
$0.6 \leq z_l < 0.8$ & 12\%$^*$  & 12\%$^*$   & 2\%   & 8\%\\
\bottomrule
\vspace{0.005in}

\end{tabular}

$^*$ Average for the \textsc{ngmix} sample.
\caption{Summary of the systematic uncertainties to be added in quadrature to the statistical error budget. $\sigma_\text{stat}$: Statistical uncertainty, as a fraction of the $b\cdot r$ values; $\sigma_\text{p-$z$}$: Photo-$z$ uncertainty on the mean of the source redshift distribution (see Sec.~\ref{subsec: photoz errors}); $\sigma_\text{m}$: Mutiplicative shear biases (see Sec.~\ref{subsec: multiplicative shear biases}); $\sigma_\text{IA}$: Intrinsic alignments (see Sec.~\ref{subsec: intrinsic alignments}).\label{tab:systematics}}
\end{table}

In all the results presented in this work, we have assumed, due to the weak lensing regime of our observations, $\kappa \ll 1$, $\left | \gamma \right | \ll 1$, that the observable reduced shear is equivalent to the shear induced by foreground mass structures, $g \approx \gamma$. Next, we provide justification for this assumption, mostly based on the work by Clampitt et al.~(2016) (henceforth Cl16) presenting the galaxy-galaxy lensing measurements around Luminous Red Galaxies (LRGs) in DES-SV along with multiple systematics tests.     

The observable reduced shear $g$ is related to the shear $\gamma$ according to Eq. (\ref{eq: reduced shear}). Since the lensing convergence $\kappa$ will always be larger for a smaller distance from the halo center, the potential differences between $g$ and $\gamma$ will be largest at the lowest scales. Cl16 estimate this difference for their smallest radial scale, $R\sim 0.1$ Mpc/$h$, and their largest estimated halo mass, $M \sim 2\times 10^{12} M_\odot / h$, and find it to be at most 0.7\%. The smallest radial scale used in this work is significantly larger than that, $R\sim 4$ Mpc/$h$, and the mean halo mass of the benchmark galaxies is expected to be smaller than the LRG sample in Cl16, since it includes galaxies from all types and luminosities. Therefore, the error we make by ignoring non-weak shear effects will be smaller than 0.7\%, and we neglect it in the analysis. Similarly, magnification can potentially affect the galaxy-galaxy lensing measurements, but it only becomes important for lenses with $\kappa$ larger than the ones in the benchmark sample (see \citealp{Mandelbaum2005} for a discussion of this effect).

\subsection{Multiplicative shear biases}\label{subsec: multiplicative shear biases}

\citet{Jarvis2015} studied the residual multiplicative shear biases for the \ngmix and \im shear catalogs, for the same redshift bins that we use in this analysis. The residual multiplicative shear biases are shown in fig. 25 of that work, and for all the redshift bins that we use are less than 1\%, except for the bin of \ngmix of $0.55 < z < 0.83$, where they reach 2\%. We decided to add 2\% of error in quadrature to the other sources of error, following the same approach as in \citet{Clampitt2016}.

\subsection{Intrinsic alignments} \label{subsec: intrinsic alignments}

Intrinsic alignments (IA) in the shapes and orientations of source galaxies can be produced by gravitational tidal fields during galaxy formation and evolution. IA can induce correlations between the source ellipticity and the lens position if the two galaxies are physically close, essentially at the same redshift. We have worked under the assumption that the observed ellipticity of a galaxy is an unbiased estimation of its shear. However, a bias can arise since there is overlap in redshift between the lens and source populations used in this analysis (see Fig.~\ref{fig: nofz}), and hence we expect a contribution from IA in the observed tangential shear measurements.

At large scales, the dominant IA contribution arises from the alignment of galaxies with the tidal field, described by the ``tidal/linear alignment model''
\citep{Catelan2001, Hirata2004, Blazek2015}. On smaller scales, non-linear contributions, including angular momentum correlations from ``tidal torquing,'' may be significant (e.g. \citealt{Lee2000}). Tidal alignment is expected to be strongest for elliptical galaxies, which are pressure supported and thus have shapes and orientations that are less affected by angular momentum. Indeed, massive elliptical galaxies exhibit stronger alignments than fainter or bluer galaxies (e.g. \citealt{Hirata2007, Mandelbaum2011}). Including the non-linear evolution of the dark matter clustering improves the linear alignment model on smaller scales, yielding the so-called "nonlinear linear alignment model" (NLA) \citep{Bridle2007}.

We estimated the contribution of IA, assuming the NLA model, for the scenarios with the most overlap between the lenses and the sources: $0.4\leq z_l<0.6$ with $0.55<z_s<1.3$ and $0.6\leq z_l<0.8$ with $0.83<z_s<1.3$. Assuming a fiducial intrinsic alignment amplitude $A = 1$, a conventional normalization chosen by Hirata \& Seljak (2004) to match ellipticity correlations in the SuperCOSMOS survey \citep{Brown2002}, a maximum fractional IA contamination on the tangential shear $\left| 1 - \gamma_\text{IA}/\gamma_t \right|$ of 4\% was obtained for these samples. Also, we found the fractional IA contamination to be nearly scale-independent, since lensing and IA are sourced by the same underlying potential. \citet{TheDarkEnergySurveyCollaboration2015} estimated the IA amplitude as $A= 2\pm 1$ for the same DES-SV source sample. This result was model dependent, and $A$ was found to be consistent with 0 for some cases. Following a conservative approach, we add an IA contamination of 8\% in quadrature to the error budget, corresponding to an uncertainty at the level of $A = 2$.

\subsection{Splitting sources in redshift} \label{subsec: redshift split sources}

For the following test, we split the source population into two separate redshift bins. Although the tangential shear measurements from two source populations with different redshift distributions $N(z)$ will have different lensing efficiencies, we can still compare the galaxy bias, since the theory predicted tangential shear also depends on the source $N(z)$. Thus, the dependency of the galaxy bias on the source redshift distribution is cancelled in case of being able to determine it precisely. Otherwise, biases in the $N(z)$ can give arise to differences between the galaxy bias from the two source bins. Hence, in some sense, this is also a \pz test.

We have separated the sources from the \ngmix shear catalog with $0.55 < z_s< 1.3$ into the two higher redshift bins used in other DES-SV weak lensing analyses (e.g. \citealt{TheDarkEnergySurveyCollaboration2015}), which are $0.55 < z_s < 0.83$, and $0.83 < z_s < 1.3$. We have performed this test on the low-$z$ lens bin from 0.2 to 0.4, to minimize the impact intrinsic alignments effects could have on the test. For the BPZ lens bin, we obtain $b\cdot r = 0.78 \pm 0.15$ for the low-$z$ source bin and $b\cdot r = 0.90 \pm 0.12$ for the high-$z$ source bin. The two results are consistent, neglecting the correlation between the two measurements.

\subsection{Observational systematic effects}

DES is a photometric survey and, as such, it is subject to changing observing conditions that may affect the galaxy catalogs and the measurements performed with them. Cr16 carried out a series of careful tests to determine and correct for any possible observational systematics in the data. In particular, they found a number of effects impacting on the detection efficiency of galaxies and hence causing density variations across the survey area. In order to study them, they used maps created from single-epoch properties potentially related to changes in the sensitivity of the survey, such as depth, seeing, airmass, etc. (see \citealp{Leistedt2015} for more details on the creation of the maps). They reported significant effects of some of these quantities on the galaxy clustering observable, especially depth and seeing variations, and they corrected for them in several ways, including using cross-correlations between the galaxy and systematics maps, and the application of a veto mask avoiding the regions most affected by these systematics. 

On the other hand, \citet{Kwan2016} studied the impact of the same systematics on galaxy-galaxy lensing, which being a cross-correlation is naturally more robust to systematic errors. They found that the effect in the galaxy-galaxy lensing observables is not significant given the statistical power 
of the observations in DES-SV. Based on these findings, we do not apply any correction from cross-correlations with systematics maps, but we do apply the veto masks in Cr16 to eliminate regions with high concentrations of these observational systematics (see Sec.~\ref{subsec: mask}). 
\section{Discussion and Comparison to previous work}
\label{sec:discussion}

\begin{table*}
\centering
\begin{tabular}{ llccc}\toprule 
Photo-$z$ code & Probe & $0.2\leq z_l < 0.4 $ & $0.4\leq z_l < 0.6 $& $0.6\leq z_l < 0.8 $\\
\midrule 

\multirow{3}{*}{BPZ} 
& g-g lensing -- This work ($b\cdot r$) & $0.87 \pm 0.11$ & $1.12 \pm 0.16$ &  $1.24 \pm 0.23$ \\
& g clustering -- \citet{Crocce2015}  ($b$)& $1.05 \pm 0.07$   & $1.23 \pm 0.05$  & $1.35 \pm 0.04$ \\
& g-CMB lensing -- \citet{Giannantonio2015} ($b\cdot r$)& $0.36 \pm 0.22$   & $0.76 \pm 0.24$  & $1.13 \pm 0.25$ \\
[0.2cm] 

\multirow{3}{*}{TPZ} 
& g-g lensing -- This work ($b\cdot r$) & $0.77 \pm 0.11$  & $1.40 \pm 0.21$ &  $1.57 \pm 0.27$ \\
& g clustering -- \citet{Crocce2015} ($b$) & $1.07 \pm 0.08$  & $1.24 \pm 0.04 $ & $ 1.34 \pm 0.05$  \\ 
& g-CMB lensing -- \citet{Giannantonio2015} ($b\cdot r$) & $0.41 \pm 0.21$   & $0.75 \pm 0.25$  & $1.25 \pm 0.25$ \\
[0.2cm] 
SkyNet & g-$\gamma$ maps -- \citet{Chang2016} ($b/r$) &  $1.12 \pm 0.19$ & $0.97 \pm 0.15 $ &  $ 1.38 \pm 0.39 $ \\

\bottomrule
\end{tabular}
\caption{ \label{tab:br results comparison} Comparison between this work's fiducial galaxy bias measurements for BPZ and TPZ lens redshift bins, galaxy clustering measurements from Cr16, galaxy-CMB lensing real space measurements from G16 and bias measurements for SkyNet lens bins from cross-correlations between galaxy density and weak lensing maps from \citet{Chang2016}.}
\end{table*}

\begin{figure}
      \hspace*{-5mm}
        \includegraphics[width=0.5\textwidth]{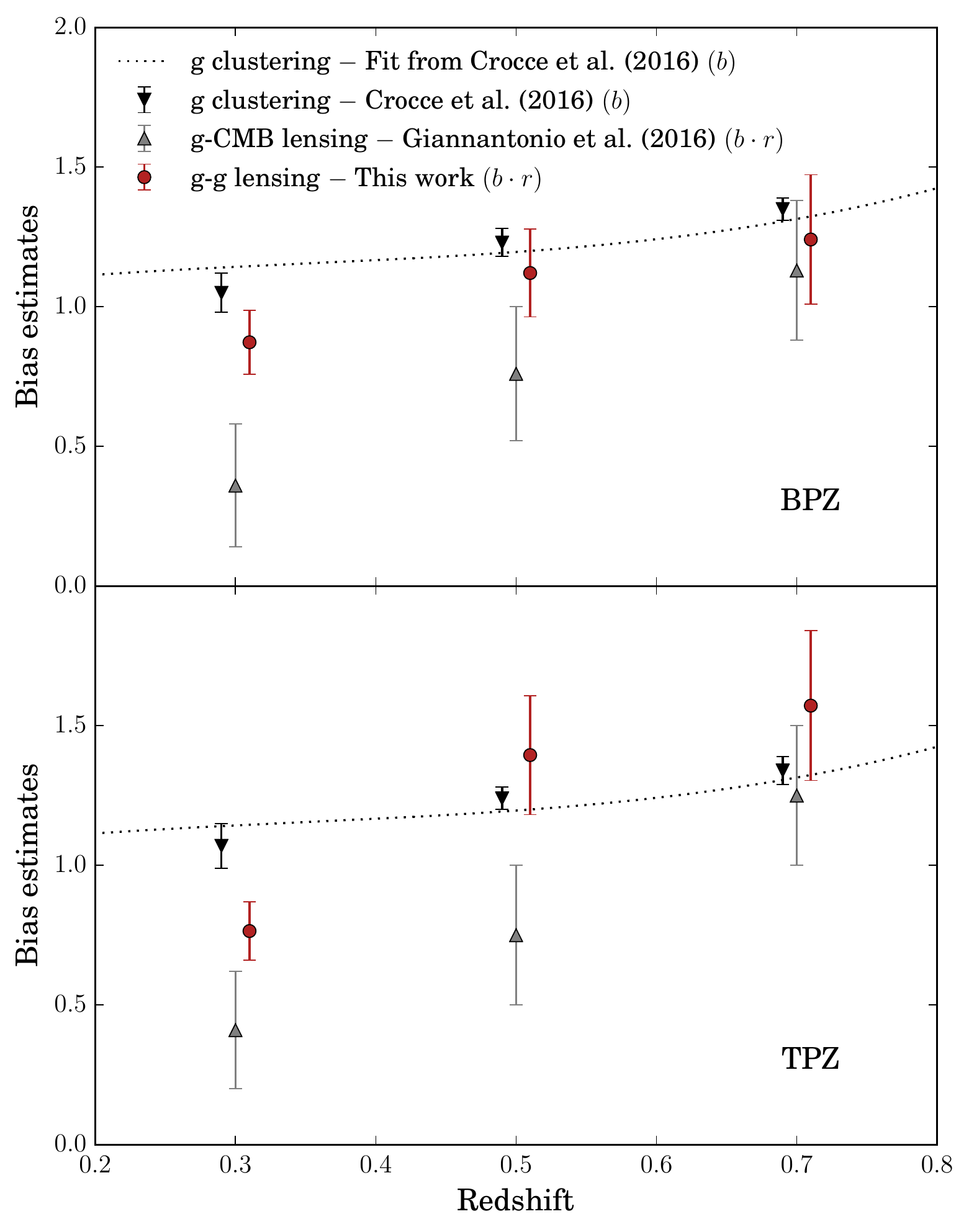}
    \caption{Fiducial galaxy bias results from this work using galaxy-galaxy lensing (g-g lensing, red points) as a function of redshift compared with previous measurements on the same Benchmark sample using galaxy clustering (g clustering, black down triangles) from Cr16 and the real space analysis results from CMB lensing ($\text{g-CMB lensing}$, gray upper triangles) from  G16. \textit{Top panel:} Lens redshift bins are defined with BPZ and the lens $N(z)$ is estimated using BPZ as well. \textit{Bottom panel:} Lens redshift bins are defined with TPZ and the lens $N(z)$ is estimated using TPZ as well. The points have been offset horizontally for clarity purposes. }
        
\label{fig: br results bpz and tpz shift}
\end{figure} 

In this work, we aim to provide another angle to the discussion of the possible tension between galaxy bias results in the DES-SV benchmark galaxy sample obtained using galaxy clustering (Cr16) and galaxy-CMB lensing correlations (G16) by adding a third probe to the discussion: galaxy-galaxy lensing. 

In Cr16, galaxy bias was measured by finding the best-fit between the galaxy angular correlation function (2PCF) and a prediction of the same function using the non-linear dark matter power spectrum. The ratio of the measurement to the unbiased theory predicion yields the square of the galaxy bias, $b^2$, from which $b$ can be directly derived.

G16 measured the cross-correlations between the galaxy density field and the lensing of the CMB, using both Planck and South Pole Telescope (SPT) maps. In that case, the comparison to theory predictions returns the galaxy bias $b$ times a factor $A_{Lens}$ which encapsulates different effects that can influence the amplitude of the CMB lensing signal. If the underlying true cosmology matches their fiducial $\Lambda$CDM model, $A_{Lens}$ should be equal to one provided the scales they use are not affected by stochasticity or non-linearities. In the general case where those can be present, their estimator yields the galaxy bias times the cross-correlation coefficient, $b \cdot r$ (cf. G16 Section 7.4). The tomographic measurements of the galaxy bias were obtained using galaxy-SPT cross-correlations, due to their higher significance. Also, both real- and harmonic-space analyses were performed, yielding consistent results. All three analyses measure the galaxy bias of the same sample of galaxies, the so-called Benchmark sample. Also, aiming for consistency, the same fiducial cosmology is assumed in all three probes: a flat $\Lambda$CDM+$\nu$ (1 massive neutrino) cosmological model based on the Planck 2013 + WMAP polarization + highL(ACT/SPT) + BAO best-fit parameters from \citet{Ade2014}.

In Fig.~\ref{fig: br results bpz and tpz shift}, as well as in Table~\ref{tab:br results comparison}, we compare our fiducial galaxy bias results with those from Cr16 and G16. In the top panel, lens redshift bins defined with BPZ are used and the $N(z)$ of the lenses is computed with BPZ for all probes, and analogously for TPZ on the bottom panel. However, different calculations of the galaxy bias cannot be directly compared since the three probes do not measure exactly the same quantity. Instead, only the galaxy clustering measurement gives a direct estimate of the galaxy bias $b$. On the other hand, both galaxy-galaxy lensing and galaxy-CMB lensing are sensitive to $b\cdot r$. Nevertheless, the cross-correlation coefficient is expected to be close to unity over the scales considered in this work, $R>4$~Mpc/$h$ (see Sec.~\ref{subsec: Range of selected scales}). 

Also, when comparing the different probes, their potential cross-covariance should be considered, since a significant covariance between them would make overall discrepancies more significant. The cross-covariance between galaxy-galaxy lensing and galaxy clustering is expected to be close to zero on small scales, where the errors are dominated by the lensing shape noise. On large scales, it is expected to be slightly higher; \citet{Mandelbaum2013} find values $\sim$ 10--15\% but compatible with zero. Similarly, \citet{Baxter2016} found that CMB-lensing and galaxy-galaxy lensing measurements on the Benchmark sample are largely uncorrelated. Henceforth, for the following discussion, we assume there is no correlation between probes. 

The different \pz bins are also covariant to some extent, given their partial overlap in redshift due to \pz errors and the use of shared sources in galaxy-galaxy lensing. Thus, since it is not easy to quantify the overall agreement of the results of the three probes across the three \pz bins, we discuss the differences between probes on a bin by bin basis. 

Then, as shown in Table~\ref{tab:br results comparison}, and neglecting the correlation between probes, our results of $b\cdot r$ are compatible within 1$\sigma$ with results of Cr16 in all redshift bins, except for the low-$z$ bin ($0.2 < z <0.4$), with a 1.3$\sigma$ difference in the BPZ case and a more significant 2.3$\sigma$ tension for TPZ. Similarly, our measurements are in moderate tension with $b\cdot r$ results of G16 (1--2$\sigma$) at low and medium redshift ($0.2 < z <0.6$), with a maximum tension of $2.1\sigma$ for the low-$z$ BPZ bin and $2.0\sigma$ for the mid-$z$ TPZ bin, while they are compatible within $1\sigma$ at higher redshift ($0.6 < z < 0.8$). However, note that the G16 results shown in Table~\ref{tab:br results comparison} and Fig.~\ref{fig: br results bpz and tpz shift} come from real space analysis. The galaxy bias results measured using harmonic space in G16 are closer to our measurements at low and medium redshift but further at higher redshift: $0.57\pm 0.25$, $0.91 \pm 0.22$, $0.68 \pm 0.28$, from the low-$z$ to the high-$z$ bin, defined with TPZ.

A fourth analysis (\citealt{Chang2016}, hereafter Ch16) also measured the galaxy bias on DES-SV data cross-correlating weak lensing shear and galaxy density maps, using the method described first in \citet{Amara2012} and later re-examined in \citet{Pujol2016}, which has the advantage that is only weakly dependent on the assumed fiducial cosmology. Ch16 measured the bias on the Benchmark sample assuming $r=1$ over the considered range of scales, and using the same lens redshift binning as the three other analyses, but adopted SkyNet to define the lens redshift bins. Thus, the lens sample slightly differs from the one used in the other three probes. In Ch16 the galaxy bias was estimated in four tomographic bins; the results obtained for the first three redshift bins can be found in Table~\ref{tab:br results comparison}, which agree at the 1--2$\sigma$ level with our measurements. 

In Fig.~\ref{fig: br results bpz and tpz shift} we observe that most of the differences between the results coming from auto-correlations and the ones coming from cross-correlations could be partially explained if $r<1$.  This would be the case if the galaxies are either stochastically or non-linearly biased, or a mixture
of both \citep{Pen1998,Simon2007}. Consider a general relation between the galaxy density contrast $\delta_g$ and the dark matter density contrast $\delta_m$:
\begin{linenomath}
\begin{equation}
\delta_g = f(\delta_m) + \epsilon\, ,
\end{equation}
\end{linenomath}
where $f$ is some function and $\epsilon$ a random variable (noise) that satisfy $\left<\epsilon \, f(\delta_m) \right> = \left<  \epsilon \,  \delta_m \right> = 0$, since $\epsilon$ is not correlated with either $f$ or $\delta_m$. If $f$ is linear ($\delta_g$ and $\delta_m$ are Gaussian random variables) and $\epsilon = 0$, we have a linear deterministic relation between $\delta_g$ and $\delta_m$: $\delta_g = b_1 \, \delta_m$. Otherwise, $f$ being a non-linear function leads to non-linear bias, and $\epsilon \neq 0 $ introduces some dispersion in the relation, usually called stochasticity. Then, $r<1$ can be generated by the presence of either non-linearities or stochasticity, or both, following from Eq. (\ref{eq: r}):
\begin{linenomath}
\begin{equation}
 r = \frac{\left< \delta_m \delta_g \right>}
 {\sqrt{\left< \delta_m^2 \right> \left< \delta_g^2\right>}} = 
 \frac{\left< \delta_m  f(\delta_m) \right>}
 {\sqrt{\left< \delta_m^2 \right> \left( \left< \left[ f(\delta_m) \right]^2\right> + \left< \epsilon^2\right> \right)}}.
 \end{equation} 
 \end{linenomath}
 
Next, we proceed to discuss the potential reasons for the possible tension between the galaxy bias estimations of the different probes, including non-linear and stochastic bias, which would both lead to $r<1$, as well as how the choice of the fiducial cosmology can affect the bias results and what is the impact of systematics effects.

\subsection{Non-linear bias} 

In Sec.~\ref{subsec: Nonlinear bias model}, we have tested the impact of using a non-linear bias modelling for scales larger than 4 Mpc/$h$, obtaining results consistent with the linear bias values. Thus, linear bias theory is currently sufficient over this range of scales given our current uncertainties.

Also, other DES-SV studies have been performed on the scale of linear bias. For instance, \citet{Kwan2016} followed a different approach to study it on the SV redMaGiC sample \citep{Rozo2015}, finding a slightly larger value of $\sim$5.5 Mpc/$h$, as expected for a red sample of galaxies. 

In G16 smaller scales were used (down to 2.4 arcmin, which approximately corresponds to 0.6 Mpc/$h$ at $z=0.3$  and to 1.2 Mpc/$h$ at $z = 0.7$), in order to extract as much signal as possible, since in their case the theoretical uncertainties due to non-linearities were much smaller than the statistical errors. Then, it is possible that non-linear bias is present over this range of scales. Refer to Sec.~\ref{subsec: Range of selected scales} for an extended discussion on the range of scales considered in this work.

\subsection{Stochasticity} 

Assuming non-linear bias can be ignored on the scales of interest and also that the fiducial cosmology (defined in Sec.~\ref{sec:method}) is fixed, it is possible to attribute the differences between the results of different probes to stochasticity. 

In this special case, G16 measured $r= 0.73 \pm 0.16$ using a novel linear growth bias-independent estimator 
-- denoted by $D_G$ in G16 -- which would imply a 1.7$\sigma$ measurement that there is some stochasticity. In G16, the measurement was extended to lower separations, where $r$ might deviate from one. Thus, this could partially explain the systematically lower results at low redshift of G16 compared to Cr16 and our measurements. 

In our case, neglecting the correlation with Cr16, in the low-$z$  bin we measure $r = 0.83 \pm 0.12$, that is, 1.3$\sigma$ away from one, for BPZ, and $r = 0.71 \pm 0.11$, 2.6$\sigma$ away from one, for TPZ, when comparing our results with those of Cr16. In the other redshift bins, the significance is much lower. Actually, since Cr16 results and this work are potentially correlated, the given confidence levels are a lower limit. 

\subsection{Fiducial cosmology dependence} 

Another possibility for the differences in the galaxy bias results is that the true cosmology does not match the fiducial cosmology, since the various probes might depend differently on the cosmological parameters. Then, even if the same fiducial cosmology is assumed (defined in Sec.~\ref{sec:method}), which is the case for Cr16, G16 and this work, this could still produce variations in the galaxy bias results. Regarding Ch16, even though a different fiducial cosmology is assumed -- the MICE cosmology \citep{Fosalba2015a, Fosalba2015b, Crocce2015a, Carretero2014} -- their approach is only weakly dependent on it. On the contrary, the galaxy bias measurements in Cr16, G16 and this work are significantly dependent on cosmology.

Particularly, the three probes are especially sensitive to $\sigma_8$ and $\Omega_m$. At large scales, if $\Omega_m$ is fixed, the galaxy bias becomes independent of scale and is hence fully degenerate with the amplitude of the matter power spectrum, $\sigma_8$. However, the dependency is different for each probe. At large scales, the galaxy clustering correlation function depends on $\sigma_8$ like $\omega^{gg}(\theta)\propto b^2 \sigma_8^2$, the tangential shear as $\gamma_t \propto b \cdot r \ \sigma_8 ^2$ and the galaxy-CMB lensing correlation function as $\omega ^{\kappa g}_{\text{CMB}}\propto b\cdot r \ \sigma_8^2$. Then, the bias from the auto-correlation depends differently on $\sigma_8$ than the bias from the cross-correlations:  $b \propto \sigma_8 ^{-1}$, $b \cdot r \propto \sigma_8 ^{-2}$. Hence, for instance, using a fiducial value of $\sigma_8$ lower than Planck's, as hinted by CFHTLenS \citep{Heymans2013}, would increase $b$, but would increase $b\cdot r$ even more, reducing the tension between probes in most of the cases. 

As an illustration to this, also involving the other cosmological parameters, G16 studied how changing the fiducial cosmology from Planck to MICE affects the galaxy bias results. MICE simulations are based on the $\Lambda$CDM cosmological parameters: $\Omega_m = 0.25$, $\Omega_{\text{DE}} = 0.75$, $\Omega_b = 0.044$, $\sigma_8 = 0.8$, $n_s = 0.95$ and $h= 0.7$. This variation of the cosmological parameters produces an increase of the galaxy bias of $\sim 4\%$ for galaxy clustering and of $\sim 21\%$ for CMB lensing (G16), for the whole redshift range $0.2 <z_l < 1.2$. For galaxy-galaxy lensing, we obtain an increase of $\sim 22\%$, for the redshift bin whose mean value is closest to the one G16 uses. Thus, the relative increases of the galaxy bias would reduce the existing tension between probes in most of the cases. 

Furthermore, we studied how the results vary performing a more plausible change in the fiducial cosmology. Using Planck 2015 + External cosmology (TT, TE, EE + lowP + Lensing + BAO + SN Ia): $\Omega_m = 0.307$, $\Omega_{\nu_{\text{mass}}} = 0.00139$, $\Omega_b = 0.0486$, $\sigma_8 = 0.816$, $n_s = 0.967$ and $h= 0.677$ \citep{PlanckCollaboration2015}, which corresponds to a $1\sigma$ variation of $\sigma_8$ with respect to the Planck 2013 value, represents an increase of $\sim 4\%$ in the bias from galaxy-galaxy lensing, not enough to account for the all the difference between probes. Further discussion on how the various cosmological parameters and models impact the bias measurements can be found in G16. 

\subsection{Photo-$\boldsymbol{z}$ errors and systematics} Another possible reason for the tension between probes is systematic error, especially in the low-$z$ bin. In \citet{Sanchez2014} it was found that the absence of $u$ band could have led to imprecise \pz measurements, particularly in the lowest redshift bin, which could potentially induce larger uncertainties in the redshift distribution of galaxies, which can affect differently each probe. For instance, auto-correlations are more sensitive to the width of the $N(z)$ distribution, while cross-correlations are more sensitive to its mean. Moreover, the shape of the CMB lensing kernel could increase the impact of \pz uncertainties at low-$z$ (see G16 for an extended discussion of how photo-$z$ can influence each probe). 

Other systematics, such as stellar contamination, can also alter the galaxy bias results in a different manner for each probe. For instance, in the case of stellar contamination, the measured galaxy clustering amplitude would be higher than otherwise, increasing the bias as well. This is already taken into account in Cr16. On the other hand, the tangential shear amplitude and the galaxy-CMB lensing cross-correlation would decrease, and so would the bias. The stellar contamination of the DES galaxy sample in the COSMOS field was found to be at most 2\% in Cr16. Although this might contribute to the observed differences in the bias, such a small contamination would produce negligible variations compared to the statistical errors. 

\vspace{6mm}
Overall, as a conclusion for the discussion presented in this section, we find no strong evidence that the cross-correlation coefficient is smaller than one, except perhaps at low redshift. In the $0.2<z_l<0.4$  bin, we measure $r = 0.83 \pm 0.12$ for BPZ, and $r = 0.71 \pm 0.11$ for TPZ, provided the differences between probes are attributed only to the cross-correlation parameter being smaller than one. Both non-linear bias and stochasticity can cause $r<1$, but, since the linear bias model is found to be a good fit for our data given the current uncertainties, our findings favor stochasticity.

Another possibility is that the differences do not have a single origin, but that they result from a combination of some of the effects presented during this discussion. Part of these potential reasons, such as a mismatch between the fiducial cosmology and the underlying true cosmology or \pz errors, while unlikely to account for the differences separately, might be able to explain them when combined. 

The DES-SV data used in this analysis represents only about 3\% of the final survey coverage. With these data, we have acquired some hints of possible causes that might have generated the differences between the results from the three probes (Cr16, G16 and this work), which will be useful for future measurements. Additional data from DES will significantly reduce the statistical uncertainties as well as allowing to probe larger scales, which will enable more precise studies of galaxy bias.

\section*{Acknowledgments}

This paper has gone through internal review by the DES collaboration. We thank the anonymous referee for her/his thorough reading of the manuscript and useful comments and suggestions that have helped improve the quality of this paper. It has been assigned DES paper id DES-2016-0173 and FermiLab preprint number PUB-16-389-AE.

Funding for the DES Projects has been provided by the U.S. Department of Energy, the U.S. National Science Foundation, the Ministry of Science and Education of Spain, the Science and Technology Facilities Council of the United Kingdom,
the Higher Education Funding Council for England, the National Center for Supercomputing Applications at the University of Illinois at Urbana-Champaign, the Kavli Institute of Cosmological Physics at the University of Chicago, the Center for Cosmology and Astro-Particle Physics at the Ohio State University, the Mitchell Institute for Fundamental Physics and Astronomy at Texas A\&M University, Financiadora de Estudos e Projetos, Funda{\c c}{\~a}o Carlos Chagas Filho de Amparo {\`a}
Pesquisa do Estado do Rio de Janeiro, Conselho Nacional de Desenvolvimento Cient{\'i}fico e Tecnol{\'o}gico and the Minist{\'e}rio da Ci{\^e}ncia, Tecnologia e Inova{\c c}{\~a}o, the Deutsche Forschungsgemeinschaft and the Collaborating Institutions in the Dark
Energy Survey.

The Collaborating Institutions are Argonne National Laboratory, the University of California at Santa Cruz, the University of Cambridge, Centro de Investigaciones Energ{\'e}ticas, Medioambientales y Tecnol{\'o}gicas-Madrid, the University of
Chicago, University College London, the DES-Brazil Consortium, the University of Edinburgh, the Eidgen{\"o}ssische Technische Hochschule (ETH) Z{\"u}rich, Fermi National Accelerator
Laboratory, the University of Illinois at Urbana-Champaign,
the Institut de Ci{\`e}ncies de l'Espai (IEEC/CSIC), the Institut
de F{\'i}sica d'Altes Energies, Lawrence Berkeley National Laboratory,
the Ludwig-Maximilians Universit{\"a}t M{\"u}nchen and
the associated Excellence Cluster Universe, the University of
Michigan, the National Optical Astronomy Observatory, the
University of Nottingham, The Ohio State University, the University
of Pennsylvania, the University of Portsmouth, SLAC
National Accelerator Laboratory, Stanford University, the University
of Sussex, Texas A\&M University, and the OzDES
Membership Consortium.

The DES data management system is supported by
the National Science Foundation under Grant Number AST-
1138766. The DES participants from Spanish institutions
are partially supported by MINECO under grants AYA2012-
39559, ESP2013-48274, FPA2013-47986, and Centro de Excelencia
Severo Ochoa SEV-2012-0234. Research leading to
these results has received funding from the European Research
Council under the European Union's Seventh Framework Programme
(FP7/2007-2013) including ERC grant agreements
240672, 291329, and 306478. Support for DG was provided by
NASA through the Einstein Fellowship Program, grant PF5-
160138.

We are grateful for the extraordinary contributions of our
CTIO colleagues and the DECam Construction, Commissioning
and Science Verification teams in achieving the excellent
instrument and telescope conditions that have made this work
possible. The success of this project also relies critically on the
expertise and dedication of the DES Data Management group.

\bibliographystyle{mn2e}
\bibliography{/Users/Judit/Dropbox/bibtex/library}

\section*{Affiliations}

\small{
$^{1}$ Institut de F\'{\i}sica d'Altes Energies (IFAE), The Barcelona Institute of Science and Technology, Campus UAB, 08193 Bellaterra (Barcelona) Spain\\
$^{2}$ Instituci\'o Catalana de Recerca i Estudis Avan\c{c}ats, E-08010 Barcelona, Spain\\
$^{3}$ Department of Physics and Astronomy, University of Pennsylvania, Philadelphia, PA 19104, USA\\
$^{4}$ Center for Cosmology and Astro-Particle Physics, The Ohio State University, Columbus, OH 43210, USA\\
$^{5}$ Department of Physics, ETH Zurich, Wolfgang-Pauli-Strasse 16, CH-8093 Zurich, Switzerland\\
$^{6}$ Jodrell Bank Center for Astrophysics, School of Physics and Astronomy, University of Manchester, Oxford Road, Manchester, M13 9PL, UK\\
$^{7}$ Institut de Ci\`encies de l'Espai, IEEC-CSIC, Campus UAB, Carrer de Can Magrans, s/n,  08193 Bellaterra, Barcelona, Spain\\
$^{8}$ Institute of Astronomy, University of Cambridge, Madingley Road, Cambridge CB3 0HA, UK\\
$^{9}$ Kavli Institute for Cosmology, University of Cambridge, Madingley Road, Cambridge CB3 0HA, UK\\
$^{10}$ Department of Physics \& Astronomy, University College London, Gower Street, London, WC1E 6BT, UK\\
$^{11}$ Institute of Cosmology \& Gravitation, University of Portsmouth, Portsmouth, PO1 3FX, UK\\
$^{12}$ Brookhaven National Laboratory, Bldg 510, Upton, NY 11973, USA\\
$^{13}$ Cerro Tololo Inter-American Observatory, National Optical Astronomy Observatory, Casilla 603, La Serena, Chile\\
$^{14}$ Department of Physics and Electronics, Rhodes University, PO Box 94, Grahamstown, 6140, South Africa\\
$^{15}$ Fermi National Accelerator Laboratory, P. O. Box 500, Batavia, IL 60510, USA\\
$^{16}$ CNRS, UMR 7095, Institut d'Astrophysique de Paris, F-75014, Paris, France\\
$^{17}$ Sorbonne Universit\'es, UPMC Univ Paris 06, UMR 7095, Institut d'Astrophysique de Paris, F-75014, Paris, France\\
$^{18}$ Kavli Institute for Particle Astrophysics \& Cosmology, P. O. Box 2450, Stanford University, Stanford, CA 94305, USA\\
$^{19}$ SLAC National Accelerator Laboratory, Menlo Park, CA 94025, USA\\
$^{20}$ Laborat\'orio Interinstitucional de e-Astronomia - LIneA, Rua Gal. Jos\'e Cristino 77, Rio de Janeiro, RJ - 20921-400, Brazil\\
$^{21}$ Observat\'orio Nacional, Rua Gal. Jos\'e Cristino 77, Rio de Janeiro, RJ - 20921-400, Brazil\\
$^{22}$ Department of Astronomy, University of Illinois, 1002 W. Green Street, Urbana, IL 61801, USA\\
$^{23}$ National Center for Supercomputing Applications, 1205 West Clark St., Urbana, IL 61801, USA\\
$^{24}$ George P. and Cynthia Woods Mitchell Institute for Fundamental Physics and Astronomy, and Department of Physics and Astronomy, Texas A\&M University, College Station, TX 77843,  USA\\
$^{25}$ Department of Physics, IIT Hyderabad, Kandi, Telangana-502285, India\\
$^{26}$ Jet Propulsion Laboratory, California Institute of Technology, 4800 Oak Grove Dr., Pasadena, CA 91109, USA\\
$^{27}$ Department of Astronomy, University of Michigan, Ann Arbor, MI 48109, USA\\
$^{28}$ Department of Physics, University of Michigan, Ann Arbor, MI 48109, USA\\
$^{29}$ Kavli Institute for Cosmological Physics, University of Chicago, Chicago, IL 60637, USA\\
$^{30}$ Department of Astronomy, University of California, Berkeley,  501 Campbell Hall, Berkeley, CA 94720, USA\\
$^{31}$ Lawrence Berkeley National Laboratory, 1 Cyclotron Road, Berkeley, CA 94720, USA\\
$^{32}$ Einstein Fellow\\
$^{33}$ Department of Physics, The Ohio State University, Columbus, OH 43210, USA\\
$^{34}$ Department of Astronomy, University of Washington, Box 351580, Seattle, WA 98195\\
$^{35}$ Australian Astronomical Observatory, North Ryde, NSW 2113, Australia\\
$^{36}$ Departamento de F\'{\i}sica Matem\'atica,  Instituto de F\'{\i}sica, Universidade de S\~ao Paulo,  CP 66318, CEP 05314-970, S\~ao Paulo, SP,  Brazil\\
$^{37}$ Department of Astrophysical Sciences, Princeton University, Peyton Hall, Princeton, NJ 08544, USA\\
$^{38}$ Department of Physics and Astronomy, Pevensey Building, University of Sussex, Brighton, BN1 9QH, UK\\
$^{39}$ Centro de Investigaciones Energ\'eticas, Medioambientales y Tecnol\'ogicas (CIEMAT), Madrid, Spain\\
$^{40}$ Universidade Federal do ABC, Centro de Ci\^encias Naturais e Humanas, Av. dos Estados, 5001, Santo Andr\'e, SP, Brazil, 09210-580\\
$^{41}$ Computer Science and Mathematics Division, Oak Ridge National Laboratory, Oak Ridge, TN 37831\\
}
\label{lastpage}

\end{document}